\documentclass[aps,prb,reprint,superscriptaddress]{revtex4-1}
\usepackage{amsmath,amssymb,graphicx,hyperref}
\DeclareMathOperator{\sgn}{sgn}

\begin{document}
\title{Classification of ground states and normal modes for phase-frustrated multicomponent superconductors}

\author{Daniel Weston}
\affiliation{Department of Theoretical Physics, KTH Royal Institute of Technology, SE-106 91 Stockholm, Sweden}

\author{Egor Babaev}
\affiliation{Department of Theoretical Physics, KTH Royal Institute of Technology, SE-106 91 Stockholm, Sweden}
\affiliation{Department of Physics, University of Massachusetts, Amherst, Massachusetts 01003, USA}

\date{2013-12-10}

\begin{abstract}
We classify ground states and normal modes for $n$-component superconductors with frustrated intercomponent Josephson couplings, focusing on $n = 4$. The results should be relevant not only to multiband superconductors, but also to Josephson-coupled multilayers and Josephson-junction arrays. It was recently discussed that three-component superconductors can break time-reversal symmetry as a consequence of phase frustration. We discuss how to classify frustrated superconductors with an arbitrary number of components. Although already for the four-component case there are a large number of different combinations of phase-locking and phase-antilocking Josephson couplings, we establish that there are a much smaller number of equivalence classes where properties of frustrated multicomponent superconductors can be mapped to each other. This classification is related to the graph-theoretical concept of Seidel switching. Numerically, we calculate ground states, normal modes, and characteristic length scales for the four-component case. We report conditions of appearance of new accidental continuous ground-state degeneracies.
\end{abstract}

\maketitle
\section{Introduction}

The generalisation of BCS theory to the multiband case with two coupled gaps was predicted in 1959,\cite{PMM.8.503, PRL.3.552} but was long widely considered a theoretical speculation. It attracted wide interest only after the 2001 discovery\cite{Nature.410.63} of superconductivity in MgB$_2$: a clear-cut example of a two-band superconductor.\cite{RPP.71.116501} In parallel, theoretical works explored to what extent two-band superconductivity differs from that in ordinary single-band BCS theory. Since the condensates in two-band superconductors are not independently conserved, they share the same broken $U(1)$ symmetry as their single-band counterparts. [For brevity, and without loss of generality, we do not distinguish between global and local $U(1)$ symmetry.] That is, although a simple two-band superconductor can under certain conditions be described by a pair of complex fields $\psi_i = |\psi_i|\mathrm{e}^{\mathrm{i}\phi_i}$ ($i=1,2$), there is a term of the form $-\eta|\psi_1||\psi_2| \cos(\phi_1-\phi_2)$, so that two-band systems attain their free-energy minimum either when the phases are locked (phase difference equal to zero), or when the phases are antilocked (phase difference equal to $\pi$).

Nevertheless, as first discussed by Leggett, the individual phases are important degrees of freedom in two-component systems. First, there exist collective excitations associated with fluctuations of the phase difference around its ground-state value.\cite{leggett,PRL.99.227002} Second, the existence of several phases can under certain conditions give rise to fractional vortices.\cite{frac}

Another qualitatively new feature which can exist in two-component superconductors is that of two coherence lengths $\xi_1$ and $\xi_2$.
(For the definition of coherence lengths in the presence of inter-component coupling in Ginzburg-Landau models, see Ref.~\onlinecite{PRL.105.067003}; for microscopic models, see Refs.~\onlinecite{PRB.84.094515,silaevGL}.) This feature can give rise to what was recently termed type-1.5 superconductivity. That is, since there exist two coherence lengths, a vortex in a two-band superconductor will typically have two cores of different sizes. In the case where $\xi_1 < \lambda < \xi_2$, where $\lambda$ is the penetration depth, there can exist thermodynamically stable vortices with long-range attractive and short-range repulsive interaction, giving rise to type-1.5 superconductivity.\cite{PRL.105.067003, PRB.72.180502, PRB.83.174509, PRB.84.134518, PRB.81.214514, PRL.102.117001, PRB.81.020506, PRB.83.020503, PRB.84.094515, silaevGL} This regime possesses an additional phase, in which there is macroscopic phase separation into domains of Meissner state and vortex state. For a recent review of type-1.5 superconductivity, see Ref.~\onlinecite{PhysicaC.479.2}.

More than half a century after two-band superconductivity was introduced, the problem of three-band superconductivity became highly relevant following the  discovery of iron-based superconductors,\cite{JACS.130.3296} which is a subject of rapidly growing interest.\cite{PhysicaC.469.313, RPP.74.124508} This raised the question if qualitatively new physics can result from the addition of a third superconducting band. It was realized that systems with three components may display phase frustration, i.e., it is not necessarily the case that the Josephson-coupling terms $-\eta_{ij}|\psi_i||\psi_j| \cos(\phi_i-\phi_j)$ can each simultaneously be minimized. Such frustration may lead to time-reversal-symmetry breaking (TRSB).\cite{EPL.87.17003, PRB.81.134522} States with TRSB break $U(1)\times \mathbb{Z}_2$ symmetry.\cite{PRB.84.134518} This phase frustration and new broken symmetry leads to interesting new physics. Recently, the scenario of such $U(1) \times \mathbb{Z}_2$ symmetry breaking in some iron-based superconductors was put on more solid ground;\cite{maiti} related states in other superconductors have also been discussed.\cite{agterberg2011} The growing number of recently discovered multiband systems has resulted in the growing opinion that multiband superconductivity is the rule rather than the exception. This has prompted investigation of multiband generalizations of various states, including the Fulde-Ferrell-Larkin-Ovchinikov state.\cite{machidanew}

In Ref.~\onlinecite{PRB.84.134518} it was discussed that phase-frustration-induced TRSB leads to new collective mixed phase-density modes, different from the phase-only Leggett mode. Normal modes were also discussed for some multiband cases in Ref.~\onlinecite{cp22}. Phase frustration can have a dramatic effect on the magnetic response of three-band superconductors.\cite{PRB.84.134518} This is because, due to frustration, the system can have large characteristic length scales associated with field variations even in the case of strong Josephson coupling. At the TRSB transition point, the length scale of one of the phase-difference modes diverges (as has also been discussed earlier in a London phase-only model\cite{PRL.108.177005}); this is a necessary consequence of the continuous transition to a ground state with higher discrete degeneracy. This divergence places systems with at least three components in contrast to two-component systems, in which increasing Josephson coupling typically diminishes disparities of the density variations.\cite{PRL.105.067003, PRB.83.174509} Also, this divergence implies that even if the system is type 2 in the TRSB state ($\lambda>\xi_i$), as one approaches the $\mathbb{Z}_2$ phase transition the divergence of one coherence length makes the system type 1.5 with $\xi_i<\lambda<\xi_j$ (provided that the phase transition is continuous\footnote{It has been argued that the transition to the TRSB phase is first order,\cite{PRB.81.134522} in which case there is no divergent length scale; however, a more recent paper\cite{maiti} supports the phase transition being second order.}). For dynamical aspects of the aforementioned mixed phase-density modes, see Refs.~\onlinecite{stanev, maiti}. Other new physics which have been discussed in connection with three-component systems are possible (meta)stable flux-carrying excitations characterized by a $\mathbb{C}P^2$ topological invariant,\cite{cp21,cp22} and fractional fluxons in long Josephson junctions\cite{PRB.86.014510} with flux quantization similar to that in fractional vortices in $[U(1)]^3$ superconductors.\cite{smiseth} Aside from that, the system can have an anomalous normal state where fluctuations destroy superconductivity, and yet the resulting normal state retains broken time-reversal symmetry.\cite{bojesen}

The above reviewed appearance of new physics in the three-component case raises the question of whether superconductors with four or more components are analogous to those with three components, or whether further new physics appears due to the additional components. Experimentally, four or more Josephson-coupled components can be realized using proximity effects in layered superconducting systems, in a similar way as in the earlier proposal to realize a three-component TRSB state in a bilayer of $s_\pm$ and $s$-wave superconductors.\cite{EPL.87.17003}

In this paper, we present a classification of superconducting states in $n$-component systems ($n \in \mathbb{N}$) with various possible frustration-inducing combinations of signs of Josephson couplings. This classification is related to the graph-theoretical concept of Seidel switching. Furthermore, we calculate ground states, normal modes, and characteristic length scales for four-component systems. In doing this, we find that the case of four and larger numbers of components is substantially richer, and allows a number of qualitatively new phenomena, as compared to the two- and three- component cases.

\section{Equivalent signatures}
\label{equivsign}

We consider multicomponent superconductors that are modelled by the London free-energy density
\begin{equation}
  f = \tfrac{1}{2}(\nabla \times \mathbf{A})^2
      + \sum_i \tfrac{1}{2} |\mathbf{D} \psi_i|^2
      - \sum_{j>i} \eta_{ij} \cos\phi_{ij}.
  \label{flondon}
\end{equation}
Here, $\mathbf{A}$ is the vector potential, $\mathbf{D} = \nabla + \mathrm{i}e\mathbf{A}$, the $\psi_i = |\psi_i| \mathrm{e}^{\mathrm{i}\phi_i}$ are complex fields representing the superconducting components, and $\phi_{ij} = \phi_i - \phi_j$. Although we here consider a London model (in which $|\psi_i| = \text{const}$), the results in this section are valid also for Ginzburg-Landau models, as well as under the inclusion of arbitrary non-phase-dependent terms in \eqref{flondon}.

We initially focus on the four-component case, although we will find results pertaining to the $n$-component case. We define the \textit{signature} of the Josephson couplings to be the tuple
\begin{equation*}
  (\sgn \eta_{12},\, \sgn \eta_{13},\, \sgn \eta_{14},\,
   \sgn \eta_{23},\, \sgn \eta_{24},\, \sgn \eta_{34}),
\end{equation*}
where $\sgn$ denotes the sign function, i.e.
\begin{equation*}
  \sgn x =
  \begin{cases}
    0 & x = 0  \\
    + & x > 0  \\
    - & x < 0. \\
  \end{cases}
\end{equation*}
The signature is of interest since each Josephson coupling sets a preferential value for a certain phase difference: if $\eta_{ij} > 0$, then $\phi_i$ and $\phi_j$ want to lock ($\phi_{ij} = 0$), whereas if $\eta_{ij} < 0$, then $\phi_i$ and $\phi_j$ want to antilock ($\phi_{ij} = \pi$). We now proceed to discuss similarities between signatures.

\begin{table}
  \caption{Representatives of the classes of strongly equivalent signatures for four components. For unfrustrated signatures, the ground-state phase configuration is given. For singly frustrated signatures, the discri\-min\-atory coupling is given.}
  \begin{ruledtabular}
  \begin{tabular}{r c c c c c c l}
    \# & $\eta_{12}$ & $\eta_{13}$ & $\eta_{14}$ &
         $\eta_{23}$ & $\eta_{24}$ & $\eta_{34}$ & Weak-equivalence class \\ \hline
     1 & $+$ & $+$ & $+$ & $+$ & $+$ & $+$
       & $\phi_1 = \phi_2 = \phi_3 = \phi_4$ \\
     2 & $+$ & $+$ & $+$ & $+$ & $+$ & $-$ & Singly frustrated ($\eta_{34}$)\\
     3 & $+$ & $+$ & $+$ & $+$ & $-$ & $-$ & Singly frustrated ($\eta_{14}$)\\
     4 & $-$ & $+$ & $+$ & $+$ & $+$ & $-$ & Multiply frustrated \\
     5 & $+$ & $+$ & $+$ & $-$ & $-$ & $-$ & Multiply frustrated \\
     6 & $+$ & $+$ & $-$ & $+$ & $-$ & $-$
       & $\phi_1 = \phi_2 = \phi_3 = \phi_4 + \pi$ \\
     7 & $+$ & $-$ & $+$ & $+$ & $-$ & $-$ & Singly frustrated ($\eta_{12}$)\\
     8 & $+$ & $+$ & $-$ & $-$ & $-$ & $-$ & Singly frustrated ($\eta_{23}$)\\
     9 & $+$ & $-$ & $-$ & $-$ & $-$ & $+$
       & $\phi_1 = \phi_2 = \phi_3 + \pi = \phi_4 + \pi$ \\
    10 & $+$ & $-$ & $-$ & $-$ & $-$ & $-$ & Singly frustrated ($\eta_{34}$)\\
    11 & $-$ & $-$ & $-$ & $-$ & $-$ & $-$ & Multiply frustrated
  \end{tabular}
  \end{ruledtabular}
  \label{signatures}
\end{table}

Under the assumption that the Josephson-coupling coefficients $\eta_{ij}$ are all non\-zero, there are $2^6 = 64$ distinct signatures. If two signatures can be mapped to each other via relabelling of the components, then they are obviously equivalent. In this case we say that the signatures are \textit{strongly equivalent}. It is easily seen that there are $11$ classes of strongly equivalent signatures; representatives of these classes are given in Table~\ref{signatures}. However, it is not necessary to study each of these equivalence classes. Instead, the signatures can be divided into three equivalence classes in such a way that it is sufficient to study a single representative of each class. We now establish this.

We define the following operators, which act on the phases and coupling coefficients, respectively:
\begin{equation*}
  P_i \colon \phi_i \mapsto \phi_i + \pi, \quad\text{and}\quad
  Q_i \colon \eta_{(ij)} \mapsto -\eta_{(ij)} \quad (\forall j \neq i).
\end{equation*}
Here, by $(ij)$ we mean the tuple obtained by sorting $i$ and $j$ in ascending order [e.g., $(12) = (21) = 12$]. In words, $P_i$ inverts the $i$th phase, and $Q_i$ changes the sign of all Josephson-coupling coefficients that involve the $i$th component. The free energy is clearly invariant under the simultaneous application of $P_i$ and $Q_i$.

Motivated by this observation, we now define another equivalence relation on the set of signatures. Two signatures are considered \textit{weakly equivalent} if one of the signatures can be obtained from the other by relabelling of the components and application of $Q_i$'s. Clearly, signatures which are strongly equivalent are also weakly equivalent. If two signatures are weakly equivalent, it is sufficient to study one of them. The reason for this is that the phase behaviour of the second signature can be obtained from the phase behaviour of the first signature via application of the appropriate $P_i$'s, and all characteristics apart from the phase behaviour are identical for the two signatures.

As a very simple example of the above, consider the cases of attractive and repulsive coupling for a two-component system (in which the phases are, respectively, locked and antilocked in the ground state). These cases are weakly equivalent since they can be mapped to each other via either of the two $Q_i$'s.

\begin{table}
  \caption{Number $N_\mathrm{J}$ of Josephson couplings, $N_\mathrm{sgn}$ of signatures, $N_\mathrm{s}$ of strong-equivalence classes (of which $N_\mathrm{su}$ unfrustrated and $N_\mathrm{sf}$ frustrated), and $N_\mathrm{w}$ of weak-equivalence classes (of which $N_\mathrm{wu}$ unfrustrated and $N_\mathrm{wf}$ frustrated) for $n$ components. We assume that the couplings are all nonzero.}
  \begin{ruledtabular}
  \begin{tabular}{r r r r r r r r r}
    $n$ & $N_\mathrm{J}$ & $N_\mathrm{sgn}$ &
    $N_\mathrm{s}$ & $N_\mathrm{su}$ & $N_\mathrm{sf}$ &
    $N_\mathrm{w}$ & $N_\mathrm{wu}$ & $N_\mathrm{wf}$
    \\ \hline
    $2$ &  $1$ &     $2$ &   $2$ & $2$ &   $0$ &  $1$ & $1$ & $0$ \\
    $3$ &  $3$ &     $8$ &   $4$ & $2$ &   $2$ &  $2$ & $1$ & $1$ \\
    $4$ &  $6$ &    $64$ &  $11$ & $3$ &   $8$ &  $3$ & $1$ & $2$ \\
    $5$ & $10$ &  $1024$ &  $34$ & $3$ &  $31$ &  $7$ & $1$ & $6$ \\
    $6$ & $15$ & $32768$ & $156$ & $4$ & $152$ & $16$ & $1$ & $15$
  \end{tabular}
  \end{ruledtabular}
  \label{equiv}
\end{table}

\subsection{Graph-theoretical approach for classification of frustrated n-component superconductors}
Our discussion so far has in no significant way been specific to the four-component case. Before moving on to the specifics of this case, we consider the general $n$-component case. In doing this, it is convenient to take a graph-theoretical approach. We let the $n$ components be represented by the (unlabeled) vertices in a graph of order $n$. The Josephson couplings are represented by edges in this graph. If a particular coupling coefficient is negative, we let the corresponding edge be blue; if a coupling coefficient is positive, we let the corresponding edge be red; if a coupling coefficient is zero, there is no corresponding edge. As long as we consider the case in which the coupling coefficients are all nonzero, we could avoid coloring the edges by letting the presence of an edge indicate one sign and the absence of an edge indicate the other sign. That the vertices are unlabeled means precisely that if two signatures are strongly equivalent, then they are represented by the same graph.

For $n$ components, there are clearly $n(n-1)/2$ possible Josephson couplings. Thus, under the assumption that the coupling coefficients are all nonzero, there are $2^{n(n-1)/2}$ signatures (removing this assumption, we get $3^{n(n-1)/2}$ signatures). The questions of how many strong-equivalence and weak-equivalence classes there are, are much more difficult to answer. However, using our graph-theoretical approach, we will make progress in this regard. Presently, we continue to assume that all of the couplings are nonzero; subsequently, we briefly consider the case in which some couplings vanish (which is relevant for multilayer or Josephson-junction-array realizations of frustrated systems).

Each strong-equivalence class corresponds to a unique complete graph on $n$ (unlabelled) vertices with edges colored red and blue. As suggested above, by removing the edges of one particular color we obtain a bijection from the set of graphs of the aforementioned type to the set of graphs on $n$ vertices (without colored edges). Thus, the number $N_\mathrm{s}(n)$ of strong-equivalence classes for $n$ components is equal to the number of graphs on $n$ (unlabelled) vertices. Although there is no known closed-form formula for this number, the corresponding enumeration problem has been solved using P\'{o}lya’s enumeration theorem.\cite{Harary.Palmer} The values of $N_\mathrm{s}(n)$ for $2 \leq n \leq 6$ are given in Table~\ref{equiv}. The sequence $N_\mathrm{s}(n)$ is Sloane's A000088.

\begin{figure}
  \includegraphics{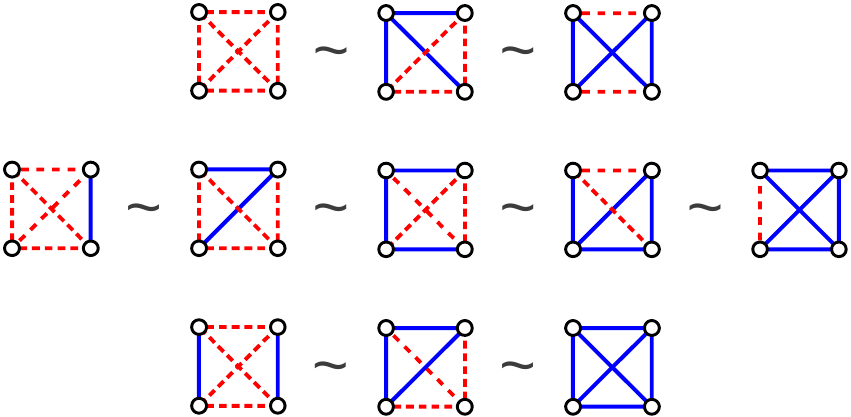}
  \caption{Switching classes of complete graphs on four vertices with edges colored red (dashed lines, attractive coupling) and blue (solid lines, repulsive coupling). The uppermost switching class corresponds to the unfrustrated signatures, the central class to the singly frustrated signatures, and the lowermost class to the multiply frustrated signatures.}
  \label{switch}
\end{figure}

Each weak-equivalence class corresponds to an equivalence class of complete graphs on $n$ (unlabelled) vertices with edges colored red and blue. The operation on such a graph corresponding to $Q_i$ is switching of the colors of all edges connected to a particular vertex. This is known as \textit{Seidel switching}, and the equivalence classes of graphs that can be transformed into each other via Seidel switching are known as \textit{switching classes} (Fig.~\ref{switch}). Thus, the number $N_\mathrm{w}(n)$ of weak-equivalence classes for $n$ components is equal to the number of switching classes of complete graphs on $n$ vertices with edges colored red and blue. As before, there is no closed-form formula for this number, but the corresponding enumeration problem has been solved.\cite{SJAM.28.876} The values of $N_\mathrm{w}(n)$ for $2 \leq n \leq 6$ are given in Table~\ref{equiv}. The sequence $N_\mathrm{w}(n)$ is Sloane's A002854.

We now illustrate the physical interpretation of the above by considering the question of how many unfrustrated strong-equivalence classes there are for $n$ components. We denote this number $N_\mathrm{su}(n)$. Without loss of generality, we assume that $\phi_1 = 0$. Clearly, each phase must have a value of $0$ or $\pi$ in order for there to be no phase frustration. This creates a partition of the phases into two sets, within which there may only be attractive couplings, and between which there may only be repulsive couplings. In terms of our graph-theoretical approach, this means that the corresponding strong-equivalence classes are represented by graphs such that the subgraph corresponding to the blue edges is a complete bipartite graph (note that this includes edgeless graphs; cf.\ Fig.~\ref{switch}). Thus, we see that $N_\mathrm{su}(n) = \operatorname{floor}(n/2 + 1)$.

Finally, we remark that there is a single unfrustrated weak-equivalence class, regardless of the number of components. This is clear since any unfrustrated phase configuration can be mapped via $P_i$'s to the configuration in which all phases are locked.

\subsection{Vanishing couplings}
We now briefly discuss the topic of systems in which some Josephson couplings vanish. We do this in the most general setting allowed by our model. The most natural goal in this context is to enumerate the weak-equivalence classes for an arbitrary number of components, without the assumption of all couplings being nonzero. In terms of our graph-theoretical approach, this means enumerating the switching classes of (not necessarily complete) graphs on $n$ vertices with edges colored red and blue. Naturally, the switching operation corresponding to a particular vertex switches the colors of edges connected to this vertex, but does not affect the presence or absence of edges. As far as we are aware, this problem has not been solved. Furthermore, we suspect that it is a very difficult problem. However, for any given moderate number of components, it is straightforward to identify the weak-equivalence classes using brute force.

\subsection{Four components}
We now apply the above to the four-component case, which we shall consider in greater detail. As mentioned above, there are three weak-equivalence classes of four-component signatures. We refer to the signatures in these classes as unfrustrated, singly frustrated, and multiply frustrated, respectively. The reasons for choosing these terms will become clear. Table~\ref{signatures} shows to which weak-equivalence class the signatures in each strong-equivalence class belong. Figure~\ref{switch} illustrates the three switching classes of graphs on four vertices which correspond to the three weak-equivalence classes of four-component signatures.

In the next section, we calculate ground states, normal modes, and characteristic length scales for $n$-component systems. Thereafter, we apply these calculations to the two weak-equivalence classes of frustrated signatures in the four-component case.

\section{Ground states and normal modes}
\label{gsnm}

We now proceed to consider the full Ginzburg-Landau free-energy density corresponding to \eqref{flondon}, i.e.,
\begin{multline}
  f = \tfrac{1}{2}(\nabla \times \mathbf{A})^2
      + \sum_i \tfrac{1}{2} |\mathbf{D} \psi_i|^2
      + \alpha_i |\psi_i|^2 + \tfrac{1}{2}\beta_i |\psi_i|^4 \\
      - \sum_{j>i} \eta_{ij} |\psi_i| |\psi_j| \cos\phi_{ij}.
  \label{fgl}
\end{multline}
It is straightforward to generalize the discussion to include terms which depend on products of $|\psi_i|$ with higher powers using the methods of Ref.~\onlinecite{PRB.83.174509}. In the following, we briefly comment also on the case of inclusion of higher Josephson harmonics. First, let us generalize some results from the case of three components\cite{PRB.84.134518} to the case of an arbitrary number of components.

Our goal in this section is twofold. First, we seek to determine the ground-state values of the $\psi_i$, i.e., the ground-state values of the densities and phases. Second, we wish to determine the normal modes of fluctuations around these ground states, as well as over what characteristic length scales such fluctuations decay. Both of these goals will be attained by expanding the relevant fields around their ground-state values, as follows:
\begin{align}
  \psi_i &= [u_i + \epsilon_i(r)]
           \exp\{\mathrm{i}[\bar{\phi}_i + \varphi_i(r)]\},\label{expansion1} \\
  \mathbf{A} &= \frac{a(r)}{r} (-\sin\theta, \cos\theta, 0)
              = \frac{a(r)}{r} \hat{\theta}. \label{expansion2}
\end{align}
Here $u_i$ and $\bar{\phi}_i$ are ground-state amplitudes and phases, respectively. Also, $r$ and $\theta$ are radial and azimuthal cylindrical coordinates, respectively. We now proceed to determine the ground states of the system in question.

\subsection{Ground states}
Inserting the field expansions \eqref{expansion1} and \eqref{expansion2} into the free-energy density \eqref{fgl} and retaining only those terms which are first order in the fluctuations, we obtain
\begin{multline}
  \sum_i 2u_i \epsilon_i(\alpha_i + \beta_i u_i^2) -
  \sum_{j>i} \eta_{ij}(u_i \epsilon_j + u_j \epsilon_i)\cos\bar{\phi}_{ij} \\
    - \eta_{ij} u_i u_j (\varphi_i - \varphi_j)\sin\bar{\phi}_{ij},
  \label{firstfluc}
\end{multline}
where $\bar{\phi}_{ij} = \bar{\phi}_i - \bar{\phi}_j$. A necessary condition for the values of $u_i$ and $\bar{\phi_i}$ to be ground-state values is that the free-energy density \eqref{fgl} is stationary with respect to fluctuations around these values. This means precisely that the prefactor of each $\epsilon_i$ and $\varphi_i$ in \eqref{firstfluc} should be zero. Requiring this, we obtain
\begin{align}
  \label{Newton1}
  0 &= \alpha_i u_i + \beta_i u_i^3
    - \tfrac{1}{2}\sum_{j \neq i} \eta_{(ij)} u_j \cos\bar{\phi}_{(ij)},\\
  \text{and}\quad
  0 &= \sum_{j \neq i}(-1)^{(i < j)} \eta_{(ij)} u_i u_j \sin\bar{\phi}_{(ij)},
  \label{Newton2}
\end{align}
for $1 \leq i \leq 4$. When we write a statement in brackets, as in $(i < j)$, we understand this to be an expression that equals one if the statement is true, and zero if the statement is false.

Unfortunately, we are unable to solve \eqref{Newton1} and \eqref{Newton2} analytically. Therefore, we determine the ground-state values $u_i$ and $\bar{\phi}_i$ numerically.

Finally, we note that it is convenient to set one of the phases to zero; this is allowed since an overall phase rotation is a pure gauge transformation. We do this in our numerical minimization, and thus the minimization is actually performed on a space with seven degrees of freedom (not eight degrees of freedom). However, in the following, we continue to work with eight degrees of freedom for reasons that will become clear.

\subsection{Length scales and normal modes}
Having considered the terms in the free-energy density \eqref{fgl} which are first order in the fluctuations, we now proceed to consider the second-order terms (which is equivalent to linearising the Ginzburg-Landau equations). In doing this, we switch to a slightly different basis; more precisely, we replace $\varphi_i$ by $\pi_i \equiv u_i \varphi_i$. The reason for this is that the so-called mass matrix, which we determine in this section, becomes symmetric in this new basis. We also introduce the notation
\begin{equation*}
  \mathbf{v} = (\epsilon_1, \dots, \epsilon_n, \pi_1, \dots, \pi_n)^\mathrm{T},
\end{equation*}
i.e.\ we collect the fluctuations of the matter fields in the vector $\mathbf{v}$.

Inserting the field expansions \eqref{expansion1} and \eqref{expansion2} into the free-energy density \eqref{fgl} and retaining only terms that are second order in the fluctuations or the gradients thereof, we obtain
\begin{equation}
  \tfrac{1}{2}(\nabla \mathbf{v})^2 +
  \tfrac{1}{2} \mathbf{v}^\mathrm{T} \mathcal{M}^2 \mathbf{v} +
  \frac{1}{2r^2} (\nabla a)^2 +
  \frac{e^2}{2r^2} \sum_i u_i^2 a^2.
  \label{quaden}
\end{equation}
We note that here the fluctuations in $\mathbf{A}$ decouple from the fluctuations in the matter fields. The matrix $\mathcal{M}^2$ is the (squared) mass matrix. Writing the corresponding terms in the free energy explicitly, we find that
\begin{widetext}
\begin{equation*}
  \tfrac{1}{2} \mathbf{v}^\mathrm{T} \mathcal{M}^2 \mathbf{v} =
  \sum_i \epsilon_i^2 \left( \alpha_i + 3\beta_i u_i^2 \right) -
  \sum_{j > i} \eta_{ij} \epsilon_i \epsilon_j \cos \bar{\phi}_{ij}
    - \eta_{ij} \bigg[ (u_i \epsilon_j + u_j \epsilon_i)
      \left( \frac{\pi_i}{u_i} - \frac{\pi_j}{u_j} \right) \sin\bar{\phi}_{ij}
      + \frac{u_i u_j}{2} \left( \frac{\pi_i}{u_i} - \frac{\pi_j}{u_j} \right)^2
      \cos\bar{\phi}_{ij} \bigg].
\end{equation*}
\end{widetext}
From this we can determine $\mathcal{M}^2$. For brevity, we introduce the notation $\bar{\eta}_{ij} = (\eta_{ij}/2) \cos\bar{\phi}_{ij}$ and $\hat{\eta}_{ij} = (\eta_{ij}/2) \sin\bar{\phi}_{ij}$. Also, we divide $\mathcal{M}^2$ into four submatrices of equal size, and extract a factor of $2$, so that
\begin{equation*}
  \mathcal{M}^2 = 2
  \begin{pmatrix}
    M_{\epsilon\epsilon} & M_{\epsilon\pi} \\
    M_{\pi\epsilon}      & M_{\pi\pi}
  \end{pmatrix}.
\end{equation*}
We are now ready to write general expressions for the above submatrices. These are
\begin{align*}
  M_{\epsilon\epsilon} &=
  \begin{pmatrix}
    0                & -\bar{\eta}_{ij} \\
    -\bar{\eta}_{ji} & 0 \\
  \end{pmatrix}
    + \operatorname{diag} \left( \alpha_i + 3\beta_i u_i^2 \right), \\
  M_{\pi\pi} &=
  \begin{pmatrix}
    0                & -\bar{\eta}_{ij} \\
    -\bar{\eta}_{ji} & 0 \\
  \end{pmatrix}
    + \operatorname{diag} \left( \frac{1}{u_i}
                               \sum_{k \neq i} u_k \bar{\eta}_{(ik)} \right),
\end{align*}
and
\begin{multline*}
  M_{\epsilon\pi} = M_{\pi\epsilon}^\mathrm{T} =
  \begin{pmatrix}
    0               & -\hat{\eta}_{ij} \\
    \hat{\eta}_{ji} & 0 \\
  \end{pmatrix} \\
    + \operatorname{diag} \left( \frac{1}{u_i}
    \sum_{k \neq i} (-1)^{(k < i)} u_k \hat{\eta}_{(ik)} \right),
\end{multline*}
where $i$ and $j$ are row and column indices, respectively.

Having written the (squared) mass matrix $\mathcal{M}^2$, let us consider its physical interpretation. First, the eigenvectors of $\mathcal{M}^2$ are the normal modes of the system. If such an eigenvector has more than one nonzero element, we say that this normal mode is mixed. We are especially interested in cases where there is both a nonzero density element and a nonzero phase element, corresponding to mixed phase-density modes. Second, the eigenvalues of $\mathcal{M}^2$ are the squared masses of the corresponding normal modes, i.e., the inverse squared characteristic length scales for the decay of small excitations of these modes.

We have seen that by diagonalizing $\mathcal{M}^2$ one can determine the normal modes and characteristic length scales of the system. Unfortunately, in general this cannot be done analytically. However, since we have included all eight (for the case of four components) degrees of freedom, we can immediately identify a normal mode, namely the gauge rotation, as well as the mass of this mode, which is zero. (Note that in an electrically charged system, this mode can acquire a mass via the Anderson-Higgs mechanism.\cite{PR.130.439}) Thus we could have limited ourselves to the seven physically relevant degrees of freedom. However, we choose not to do this, since the aforementioned knowledge about the eigenvectors and eigenvalues of $\mathcal{M}^2$ provides a useful way to check our numerical results.

\subsection{Massless modes}
\label{masslessmodes}
It is a rather general feature of frustrated multicomponent superconductors that they may undergo continuous transitions whereby discrete ground-state degeneracy arises. Examples of this are the aforementioned TRSB transitions in three-component superconductors, as well as other such transitions which are studied below. Also, such transitions may appear in phase-only models.\cite{PRL.108.177005} These transitions are quite generally accompanied by the presence of at least one massless normal mode, i.e., by the divergence of a characteristic length scale for the decay of such a mode.

Indeed, consider a potential $U(\mathbf{x},\alpha)$, which depends on the (generalized) coordinate vector $\mathbf{x} \in \mathbb{R}^n$ and the parameter $\alpha \in \mathbb{R}$. In our case, $U$ is the Ginzburg-Landau potential in \eqref{fgl}, and $\mathbf{x}$ is a vector of densities and phases (or some other parametrization of the state space). The parameter $\alpha$ can in our case have several meanings; for example, $\alpha$ could be a Josephson-coupling coefficient. We choose the above notation in order to emphasize the generality of the material in this section.

\begin{figure}
  \includegraphics{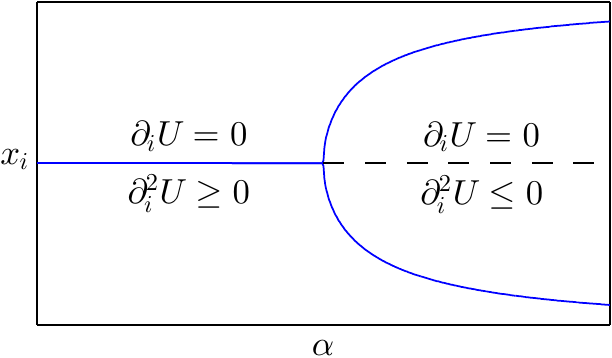}
  \caption{The ground-state value of the (generalized) coordinate $x_i$ as a (potentially multivalued) function of the parameter $\alpha$. By assumed continuity, we have that $\partial_i^2 U = 0$ at the critical point. Thus there exists a massless mode at this point.}
  \label{potential}
\end{figure}

Assume that $U(\mathbf{x},\alpha) \in C^2(\mathbb{R}^{n+1})$. Assume further that, as $\alpha$ is varied, the system undergoes a continuous transition whereby a ground state splits into two degenerate ground states. This situation is illustrated in Fig.~\ref{potential}. (The argument applies more generally to any situation where a local minimum is continuously transformed into several local minima, but this is the main case of interest to us.) We choose our coordinate system so that the Hessian of $U(\mathbf{x})$ is diagonal at the transition point. This is possible since the Hessian is symmetric, and thus diagonalizable (by an orthogonal transformation). Also, we observe that this choice of coordinates is such that each coordinate corresponds to a normal mode at the critical point $\alpha = \alpha_\mathrm{c}$ at which the transition takes place.

Choose a coordinate $x_i$ in which there is discrete degeneracy for $\alpha > \alpha_\mathrm{c}$, and consider the curve of ground states in $(x_i,\alpha)$-space which is illustrated schematically in Fig.~\ref{potential}. Obviously, each ground-state point (blue curve) is such that $\partial_i U = 0$ and $\partial_i^2 U \geq 0$ ($\partial_i$ denotes differentiation with respect to $x_i$). Now, fix a value $\alpha > \alpha_\mathrm{c}$, and consider how $\partial_i U$ varies as $x_i$ is varied: in other words, as one moves along a vertical line in the right hand side of Fig.~\ref{potential}. Immediately above the lower ground-state curve $\partial_i U > 0$, since $\partial_i U = 0$ and $\partial_i^2 U \geq 0$ on the curve and $\partial_i U \neq 0$ immediately above the curve (lest points immediately above the curve also be ground states). Similarly, $\partial_i U < 0$ immediately below the upper curve. Hence there is a point between the curves such that $\partial_i U = 0$ and $\partial_i^2 U \leq 0$. Since this holds arbitrarily close to the critical point, there is some curve (dashed line in Fig.~\ref{potential}) that emanates from the critical point and along which $\partial_i U = 0$ and $\partial_i^2 U \leq 0$. By the assumed continuity of $\partial_i^2 U$, we have that $\partial_i^2 U = 0$ at the critical point. This implies the existence of a massless mode at this point. Finally, we note that since each of the coordinates we use correspond to a normal mode at the critical point, there will be a massless mode for each coordinate in which discrete degeneracy arises at this point.

\section{Singly frustrated signatures}

In this section, and the next, we apply the results of the previous section to the four-component case. Recall that we named the two frustrated weak-equivalence classes of four-component signatures singly frustrated and multiply frustrated. The \textit{singly frustrated signatures} are the frustrated signatures for which there exists a phase configuration in which only one Josephson coupling is frustrated. We call such a phase configuration a \textit{singly frustrated phase configuration}, and we call other frustrated phase configurations \textit{multiply frustrated phase configurations}. For each singly frustrated signature there is a unique singly frustrated phase configuration [up to the overall $U(1)$ symmetry]. Thus there is, for each singly frustrated signature, a unique coupling which is frustrated in the singly frustrated phase configuration. We call these couplings the \textit{discri\-min\-atory couplings} (Table~\ref{signatures}). We now consider the effects of varying a discriminatory coupling.

\subsection{Discriminatory couplings}
If, for a given singly frustrated signature, the discri\-min\-atory coupling is sufficiently weak, then the phases will assume the singly frustrated configuration (at least, the singly frustrated configuration will be the ground-state configuration). Conversely, if the discri\-min\-atory coupling is sufficiently strong, then the phases will assume a multiply frustrated configuration. At the transition between singly frustrated and multiply frustrated phase configurations, there is a massless mode (apart from the mode corresponding to the gauge symmetry). This is an example of the general fact that continuous phase transitions are accompanied by massless modes. At the transition, time-reversal symmetry is broken, and the spontaneously broken symmetry changes from $U(1)$ to $U(1) \times \mathbb{Z}_2$. The transformation corresponding to the $\mathbb{Z}_2$ symmetry is complex conjugation of the $\psi_i$. Furthermore, for strong discriminatory couplings the corresponding phases may lock (for attractive couplings) or antilock (for repulsive couplings) leading to a second transition, this time from $U(1) \times \mathbb{Z}_2$ back to $U(1)$. In cases of TRSB, the normal modes are generally mixed. Evidently, the case of singly frustrated signatures is largely analogous to the case of frustrated three-component signatures. This is true despite the fact that for three components there is no discriminatory coupling.

\begin{figure}
  \includegraphics{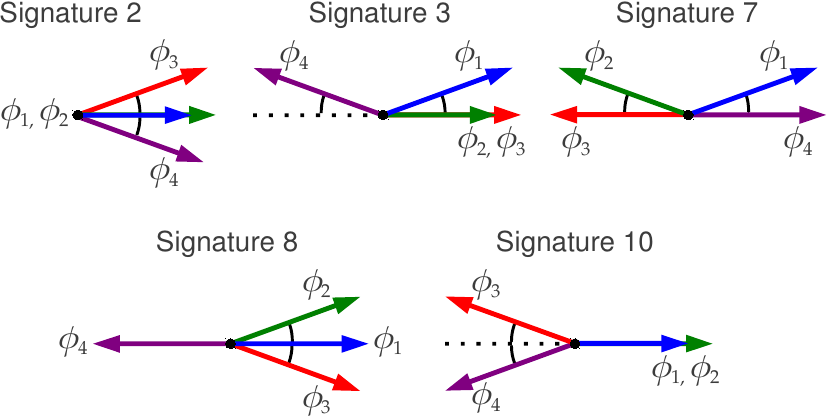}
  \caption{Examples of ground-state phase configurations for singly frustrated signatures with discriminatory couplings somewhat stronger than the critical values. There is $\mathbb{Z}_2$ degeneracy corresponding to complex conjugation of the $\psi_i$. The corresponding singly frustrated phase configurations are obtained by reducing the marked angles to zero.}
  \label{singfrust}
\end{figure}

\begin{figure*}
  \includegraphics{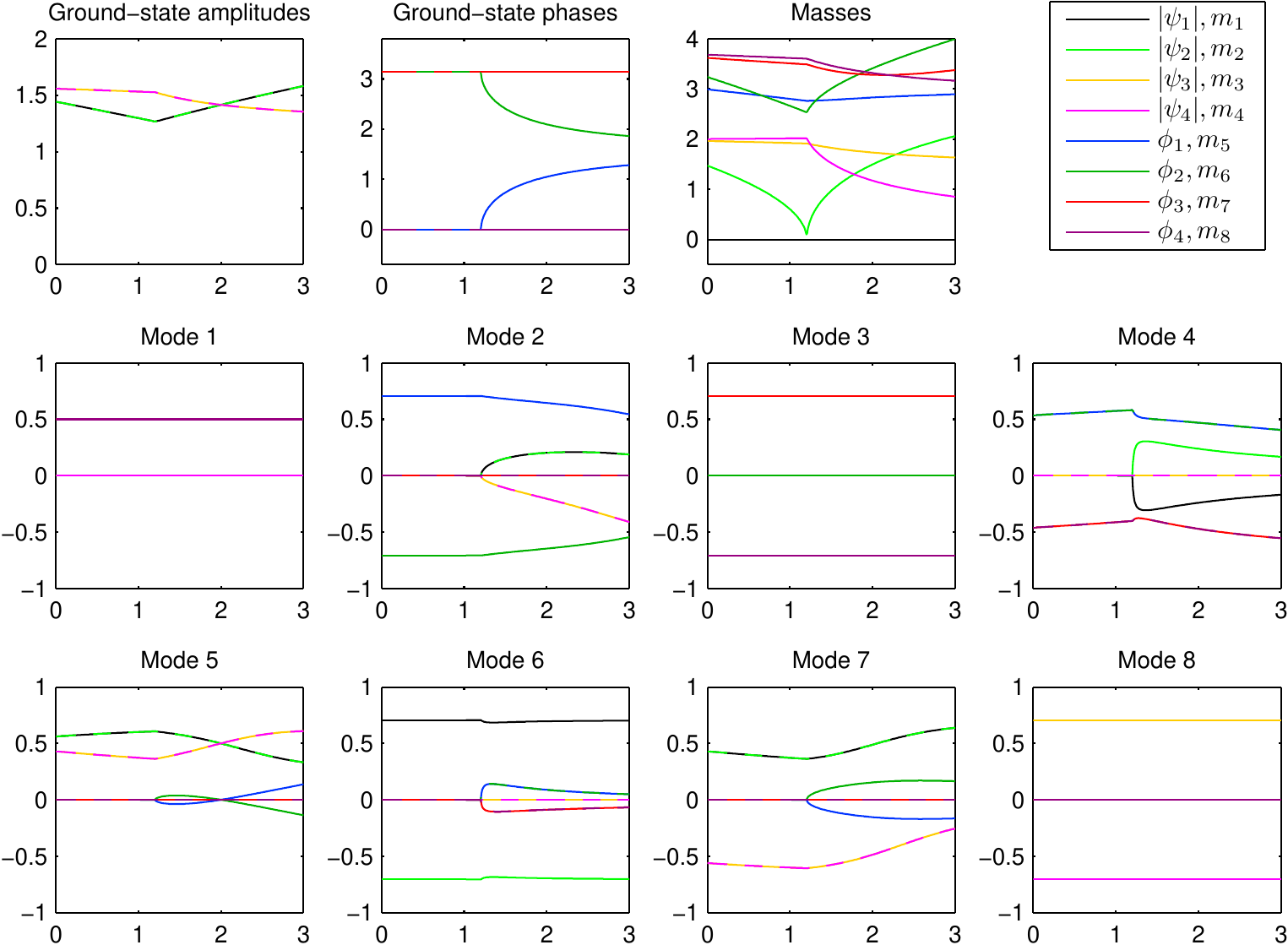}
  \caption{Ground states (cf.\ Fig.~\ref{singfrust}), (inverse) length scales, and normal modes for signature 7 with $\alpha_i = -1$, $\beta_i = 1$, and $|\eta_{ij}| = 1$  for $ij \neq 12$. $\eta_{12}$ is plotted on the $x$-axes. Note that there is a second massless mode at the critical point $\eta_{12} = \eta_{12}^\mathrm{c} = 1.21$. Also, several of the modes are mixed phase-density modes when $\eta_{12} > \eta_{12}^\mathrm{c}$, whereas none of the modes are mixed phase-density modes when $\eta_{12} < \eta_{12}^\mathrm{c}$.}
  \label{plot7}
\end{figure*}

We now consider a specific example of a singly frustrated signature. Arbitrarily, and without loss of generality, we choose signature 7. We use the free-energy parameters in \eqref{fgl} given by $\alpha_i = -1$, $\beta_i = 1$, and $|\eta_{ij}| = 1$ except that we vary the coefficient of the discri\-min\-atory coupling. The singly frustrated phase configuration for signature 7 is
\begin{equation*}
  \phi_1 = \phi_2 + \pi = \phi_3  + \pi = \phi_4.
\end{equation*}
This is the ground-state phase configuration for values of the discri\-min\-atory coupling coefficient $\eta_{12}$ smaller than the critical value $\eta_{12}^\mathrm{c} = 1.21$. For $\eta_{12} > \eta_{12}^\mathrm{c}$, the phases $\phi_1$ and $\phi_2$ approach each other by breaking their locking with $\phi_4$ and $\phi_3$, respectively, by equal and opposite amounts (Fig.~\ref{singfrust}). The ground states, normal modes and characteristic length scales for the present parameters with $0 \leq \eta_{12} \leq 3$ are shown in Fig.~\ref{plot7}. As we observed in the previous section, the normal modes are given by the eigenvectors of the (squared) mass matrix, and the characteristic length scales are given by the corresponding eigenvalues. We note that here, in contrast to the case considered in Ref.~\onlinecite{PRB.84.134518}, there is a phase-only mode (mode 3) also in the TRSB regime. Thus, such modes are possible, even though in the case of TRSB the modes are typically mixed phase-density modes. Finally, the corresponding plots for other singly frustrated signatures are identical (up to relabelling of components), except that the ground-state phase configurations are different.

\subsection{Higher ground-state degeneracy}
\label{singdisc}
An interesting question is whether there exist ground states with higher than twofold degeneracy [we here ignore the overall $U(1)$ symmetry, which is always present]. We now go some way towards answering this question for singly frustrated signatures. One way in which higher discrete degeneracy may arise is through equivalence of components: If two (or more) components are equivalent, and if the ground-state values of the corresponding fields are not equal, then exchanging the values of these fields will map a given ground state to a distinct but equivalent state. We begin by noting that for singly frustrated signatures, no more than two components can be equivalent. To see this, note that neither of the two phases coupled by the discriminatory coupling (for signature 2: $\phi_3$ and $\phi_4$ in Fig.~\ref{singfrust}) can be equivalent to either of the other two phases.

Furthermore, apart from equivalence of components, one could imagine that higher degeneracy could arise through what may reasonably be called ground-state equivalence of phases. By this we mean the following: Consider a given ground state, and in particular components $i$ and $j$. If $\eta_{(ik)} |\psi_i| = \eta_{(jk)} |\psi_j|$ ($i \neq k \neq j$) for the ground-state values of $|\psi_i|$ and $|\psi_j|$, then exchanging the values of $\phi_i$ and $\phi_j$ will have no effect on the potential energy. Thus, if it is also the case that $\phi_i \neq \phi_j$ in a ground state, then there is corresponding ground-state degeneracy. As far as we are aware, each component always has a unique ground-state value of the density. Assuming this, we have that equivalence of components is a special case of ground-state equivalence of phases.

Let $\phi_i$ and $\phi_j$ be equivalent in a ground state. Note that upon application of $P_i$ and $Q_i$, $\phi_i$ and $\phi_j$ need no longer be equivalent in the above sense. Nonetheless, any ground-state degeneracy is unaffected by this transformation. We understand that $\phi_i$ and $\phi_j$ \emph{are} in fact still equivalent in some weaker sense. For simplicity, we consider signature 2, for which this question does not arise.

Now, if $\phi_1$ and $\phi_2$ are equivalent in a ground state, this can only give rise to higher degeneracy in the aforementioned way if $\phi_1 \neq \phi_2$. Somewhat less obvious is the fact hat if $\phi_3$ and $\phi_4$ are equivalent in a ground state, then it is again necessary to have $\phi_1 \neq \phi_2$ in order for this to yield higher degeneracy. The reason for this is that if $\phi_3$ and $\phi_4$ are equivalent, and $\phi_1 = \phi_2$ in the ground state, then the degeneracy corresponding to exchange of $\phi_3$ and $\phi_4$ coincides with the degeneracy corresponding to complex conjugation.

We now show that if two phases are equivalent in a ground state, then $\phi_1 = \phi_2$ in this state, whence higher degeneracy cannot arise in the aforementioned way for singly frustrated signatures. However, we will find that such degeneracy \emph{can} occur for multiply frustrated signatures. We now in turn consider the cases of $\phi_1$ and $\phi_2$ being equivalent, and of $\phi_3$ and $\phi_4$ being equivalent.

\begin{figure}
  \centering
  \includegraphics{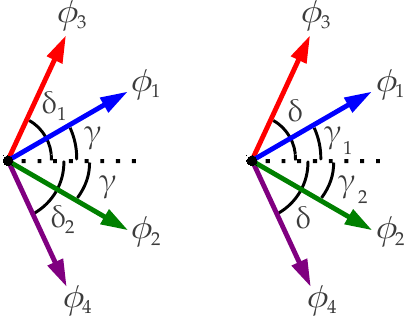}
  \caption{Parametrizations of the three relevant degrees of freedom in the phases.}
  \label{phaseparam}
\end{figure}

We parametrize the three relevant degrees of freedom in the phases as shown in the left half of Fig.~\ref{phaseparam}. For brevity, we introduce the notation $\tilde{\eta}_{ij} = -\eta_{ij} |\psi_i| |\psi_j|$. The part of the potential energy which depends on $\gamma$ is
\begin{multline*}
  F_\gamma =
  \tilde{\eta}_{13} \cos(\delta_1 - \gamma) +
  \tilde{\eta}_{23} \cos(\delta_1 + \gamma)\\ +
  \tilde{\eta}_{24} \cos(\delta_2 - \gamma) +
  \tilde{\eta}_{14} \cos(\delta_2 + \gamma) +
  \tilde{\eta}_{12} \cos2\gamma.
\end{multline*}
By the assumed equivalence of $\phi_1$ and $\phi_2$, we have that $\tilde{\eta}_{13} = \tilde{\eta}_{23} =: \tilde{\eta}_{1}$ and $\tilde{\eta}_{14} = \tilde{\eta}_{24} =: \tilde{\eta}_{2}$. Thus, using a trigonometric identity, we can write $F_\gamma$ as
\begin{equation}
  F_\gamma = 2 (\tilde{\eta}_{1} \cos\delta_1 +
                \tilde{\eta}_{2} \cos\delta_2) \cos\gamma
             + \tilde{\eta}_{12} \cos 2\gamma.
  \label{gammaen1}
\end{equation}
If $\tilde{\eta}_{12} < 0$, as we assume, then the last term in \eqref{gammaen1} is minimized for $\gamma = 0$ and $\gamma = \pi$, both corresponding to $\phi_1 = \phi_2$. Clearly, one of these values of $\gamma$ also minimizes the first term in \eqref{gammaen1}, and thus we have that $\phi_1 = \phi_2$ in the ground state.

We now parametrize the three relevant degrees of freedom in the phases as shown in the right half of Fig.~\ref{phaseparam}. The part of the potential energy which depends on $\gamma_1$ and $\gamma_2$ is
\begin{multline*}
  F_\gamma =
  \tilde{\eta}_{13} \cos(\delta - \gamma_1) +
  \tilde{\eta}_{23} \cos(\delta + \gamma_2) +
  \tilde{\eta}_{24} \cos(\delta - \gamma_2)\\ +
  \tilde{\eta}_{14} \cos(\delta + \gamma_1) +
  \tilde{\eta}_{12} \cos(\gamma_1 + \gamma_2).
\end{multline*}
By the assumed equivalence of $\phi_3$ and $\phi_4$, we have that $\tilde{\eta}_{13} = \tilde{\eta}_{14} =: \tilde{\eta}_{1}$ and $\tilde{\eta}_{23} = \tilde{\eta}_{24} =: \tilde{\eta}_{2}$. Thus we can write $F_\gamma$ as
\begin{equation}
  F_\gamma = 2(\tilde{\eta}_{1} \cos\gamma_1
             + \tilde{\eta}_{2} \cos\gamma_2)\cos\delta
             + \tilde{\eta}_{12} \cos(\gamma_1 + \gamma_2).
  \label{gammaen2}
\end{equation}
Without loss of generality, we assume that $\cos\delta \geq 0$. Since we also have that $\tilde{\eta}_{1} < 0$ and $\tilde{\eta}_{2} < 0$, we can conclude that the first term in \eqref{gammaen2} is minimized by $\gamma_1 = \gamma_2 = 0$. Also, the second term in \eqref{gammaen2} is clearly minimized precisely if $\gamma_1 + \gamma_2 = 0$. Thus, we again have that $\phi_1 = \phi_2$ in the ground state.

\section{Multiply frustrated signatures}

The \textit{multiply frustrated signatures} are the frustrated signatures for which more than one of the Josephson couplings are frustrated, regardless of the phase configuration. For such signatures, the ground-state degeneracy can be greater than for singly frustrated four-component signatures, or for fewer than four components. First, for some values of the free-energy parameters, there exists continuous ground-state degeneracy corresponding to rotation of a pair of phases relative to the other two phases. Such rotations can occur when the phases are pairwise equivalent in the aforementioned sense, and occur despite all phases being coupled. Second, for certain other values of the free-energy parameters, there can exist other types of additional continuous ground-state degeneracy. This can occur when two or three phases are equivalent.

\subsection{Energetically free phase rotations}
\begin{figure}
  \includegraphics{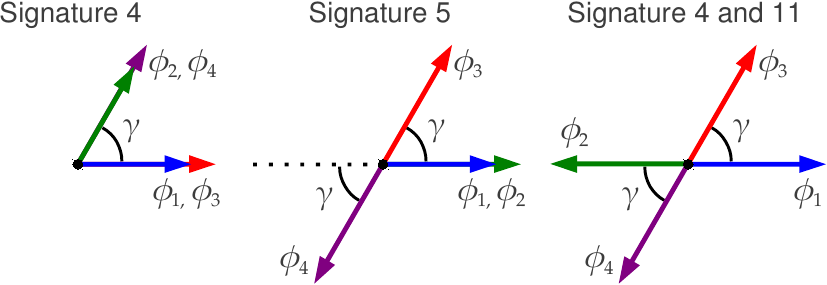}
  \caption{Examples of ground-state phase configurations for multiply frustrated signatures with complete intercomponent symmetry. The phases are pairwise locked or antilocked. The phase rotations corresponding to alteration of the angle $\gamma$ are energetically free.}
  \label{multfrust}
\end{figure}

We now say something about what parameter values give rise to energetically free phase rotations of the aforementioned type. In doing this, and in the remainder of this section, we choose to consider signature 11 (Table~\ref{signatures}). For this signature, energetically free phase rotations may exist in cases where the phases are pairwise antilocked in the ground state. Assume that $\phi_1 = \phi_2 + \pi$ and $\phi_3 = \phi_4 + \pi$, and let $\gamma = \phi_3 - \phi_1$. This situation is illustrated in the rightmost part of Fig.~\ref{multfrust}. In order for this to be a ground-state configuration with $0 \neq \gamma \neq \pi$, it is necessary that
\begin{equation}
  \tilde{\eta}_{13} = \tilde{\eta}_{14} =
  \tilde{\eta}_{23} = \tilde{\eta}_{24},
  \label{cond1}
\end{equation}
where $\tilde{\eta}_{ij} = -\eta_{ij} |\psi_i| |\psi_j|$. To see that these equalities are necessary, note that if they do not hold, one of the phases will be subject to a net force, which will tend to alter the phase configuration. Apart from being necessary ground-state conditions, the conditions in \eqref{cond1} are sufficient for the $\gamma$-rotation to be energetically free: the portion of the potential energy which depends on $\gamma$ is
\begin{multline}
  \tilde{\eta}_{13} \cos \gamma +
  \tilde{\eta}_{23} \cos(\pi - \gamma) +
  \tilde{\eta}_{24} \cos \gamma +
  \tilde{\eta}_{14} \cos(\pi - \gamma)\\
  = (\tilde{\eta}_{13} + \tilde{\eta}_{24}
   - \tilde{\eta}_{23} - \tilde{\eta}_{14})\cos\gamma.
  \label{phaserot}
\end{multline}
Conditions equivalent to those in \eqref{cond1} can of course be given for the other multiply frustrated signatures. Note that the conditions in \eqref{cond1} are equivalent to the condition of pairwise ground-state equivalence of phases.

\begin{figure}
  \includegraphics{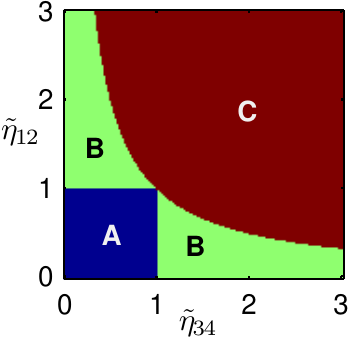}
  \includegraphics[trim = 1 1 1 1, clip]{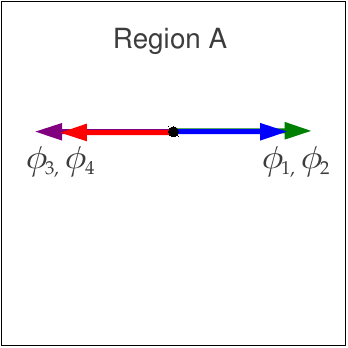}
  \includegraphics[trim = 1 1 1 1, clip]{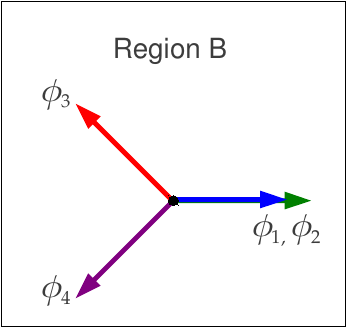}
  \includegraphics[trim = 1 1 1 1, clip]{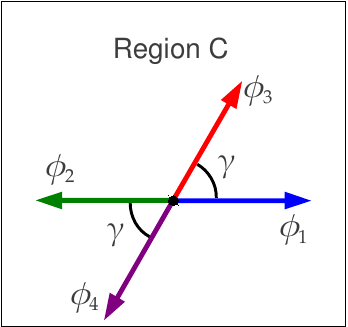}
  \caption{Classification of ground states for signature 11 under the assumption of pairwise ground-state equivalence of phases. In region A there is no TRSB, in region B there is TRSB, and in region C there are energetically free phase rotations and thus degeneracy between states with and without TRSB. Representative ground states are shown for the three regions. We set $\tilde{\eta}_{13} = \tilde{\eta}_{14} = \tilde{\eta}_{23} = \tilde{\eta}_{24} = 1$. Along the curve $\tilde{\eta}_{12} \tilde{\eta}_{34} = 1$, the situation is more complicated than this classification suggests (Fig.~\ref{special2}).}
  \label{special1}
\end{figure}

\begin{figure*}
  \includegraphics{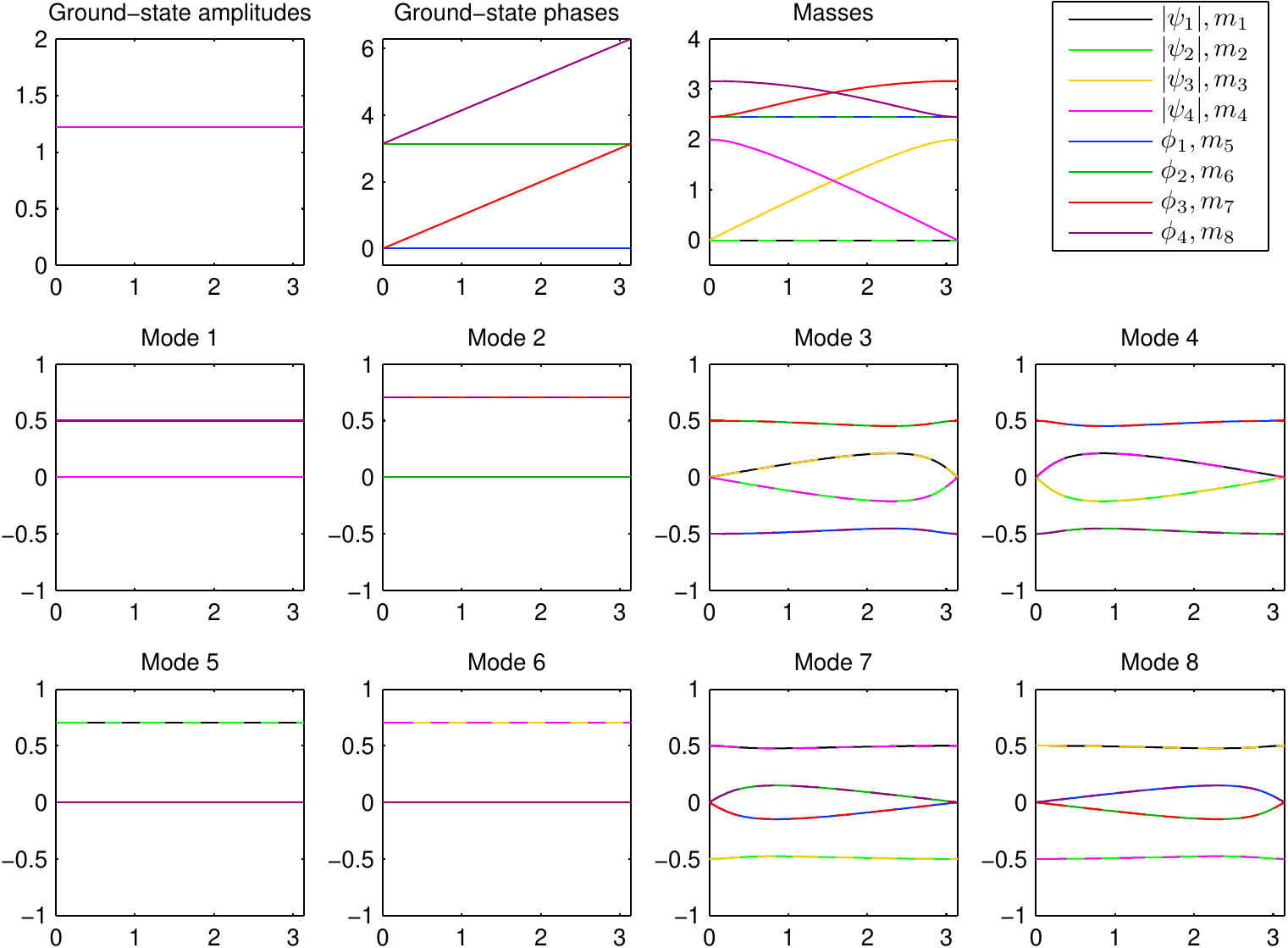}
  \caption{Ground states, (inverse) length scales and normal modes for signature 11 with $\alpha_i = -1$, $\beta_i = 1$ and $\eta_{ij} = -1$. The rotational angle $\gamma$ (Fig.~\ref{multfrust}) is plotted on the $x$-axes. Note that there is a third massless mode at the points $\gamma = 0$ and $\gamma = \pi$. Also, these points are the only points for which there are no mixed phase-density modes.}
  \label{plot11_1}
\end{figure*}

In order for the phase configurations parametrized by $\gamma$ to correspond to ground states, it is necessary that the couplings between the paired phases be sufficiently strong, so that antilocking is maintained. In fact, we find numerically that the condition is
\begin{equation}
  \tilde{\eta}_{12} \tilde{\eta}_{34} \geq \tilde{\eta}^2,
  \label{cond2}
\end{equation}
where $\tilde{\eta} := \tilde{\eta}_{13} = \tilde{\eta}_{14} = \tilde{\eta}_{23} = \tilde{\eta}_{24}$. Hence, we have found that the necessary and sufficient conditions for additional ground-state degeneracy in the form of energetically free phase rotations are \eqref{cond1} and \eqref{cond2}. Figure~\ref{special1} classifies the possible ground states under the assumption of \eqref{cond1}. (Without loss of generality, we set $\tilde{\eta} = 1$.) In cases for which $\tilde{\eta}_{12} \leq 1$ and $\tilde{\eta}_{34} \leq 1$ (region A), the ground state only breaks $U(1)$ symmetry. The energetically free phase rotations correspond to region C. The remaining possibilities (region B) give rise to ground states that break $U(1) \times \mathbb{Z}_2$ symmetry.

We now consider the case of complete symmetry between the four components, which of course leads to fulfilment of \eqref{cond1} and \eqref{cond2} in the ground state. For illustrative purposes, we choose the parameter values $\alpha_i = -1$, $\beta_i = 1$ and $\eta_{ij} = -1$. For this completely symmetric case, the ground-state phase configurations are precisely those for which the phases are pairwise antilocked. Thus, there is continuous ground-state degeneracy. Figure~\ref{multfrust} illustrates the ground-state phase configurations for this signature, as well as the corresponding phase configurations for the other two multiply frustrated signatures. As we expect, the ground states for the different signatures can be mapped to each other via relabelling of the components and inversion of the phases (application of $P_i$'s).

\begin{figure*}
  \includegraphics{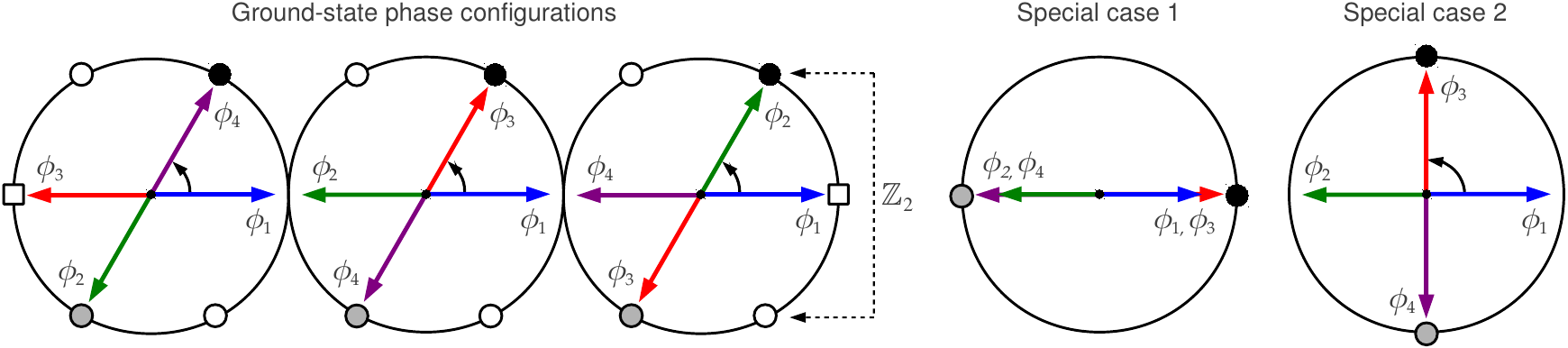}
  \caption{Illustration of ground-state phase configurations for signature 11 with complete intercomponent symmetry. We assume that $\phi_1 = 0$. There are three possible pairwise antilockings, corresponding to the three circles. In each circle, the displayed phase configuration corresponds to the black dot; other equivalent possibilities are given by the other dots. Points marked with squares coincide. Permutation of the three unfixed phases corresponds either to moving \emph{within} the set of black and grey (dark) dots, or to moving \emph{within} the set of white dots. Complex conjugation corresponds to moving \emph{between} the set of dark dots and the set of white dots. There are two types of special cases in which the number of equivalent states is smaller than in the typical case.}
  \label{symmetry}
\end{figure*}

We note that although rotation by the angle $\gamma$ ($\gamma$-rotation, Fig.~\ref{multfrust}) does not alter the potential energy, such rotation does alter the normal modes and the corresponding length scales. This is clear from Fig.~\ref{plot11_1}, which displays the ground states, normal modes, and characteristic length scales for the system in question. Also noteworthy is the fact that $\gamma$-rotation does not in itself lead to exploration of the entire family of ground states, since the pairwise antilocking of phases can occur in three ways, and rotation by $\gamma$ does not change the antilocking. This is illustrated schematically in Fig.~\ref{symmetry}.

Now, choose an angle $\gamma$ such that $0 < \gamma < \pi/2$, and consider the corresponding phase configuration in Fig.~\ref{multfrust}. From Fig.~\ref{plot11_1} it is clear that no two values of $\gamma$ in the aforementioned range give rise to equivalent normal modes. Due to the complete intercomponent symmetry, any permutation of $\phi_2$, $\phi_3$ and $\phi_4$ will give an equivalent state (Fig.~\ref{symmetry}). Furthermore, complex conjugation of each $\psi_i$ also yields an equivalent state. Thus, for a given ground state, there are typically twelve ground states that are equivalent to this state. However, there are special cases for which the number of equivalent states is three (special case 1 in Fig.~\ref{symmetry}) or six (special case 2 in Fig.~\ref{symmetry}).

Consider the ground states for which antilocked phase pairs are parallel. As can be seen from Fig.~\ref{plot11_1}, there is a third massless mode in these states (the first two massless modes being gauge rotation and $\gamma$-rotation). Naturally, this third mode corresponds to the other possible way of maintaining pairwise antilocking. The occurrence of this mode is another example of the general situation discussed in Sec.~\ref{masslessmodes}. One might object that in this case there is no parameter actually modifying the potential. Nevertheless, we can simply replace $\alpha$ by $\gamma$ without invalidating the argument.

\begin{figure*}
  \includegraphics{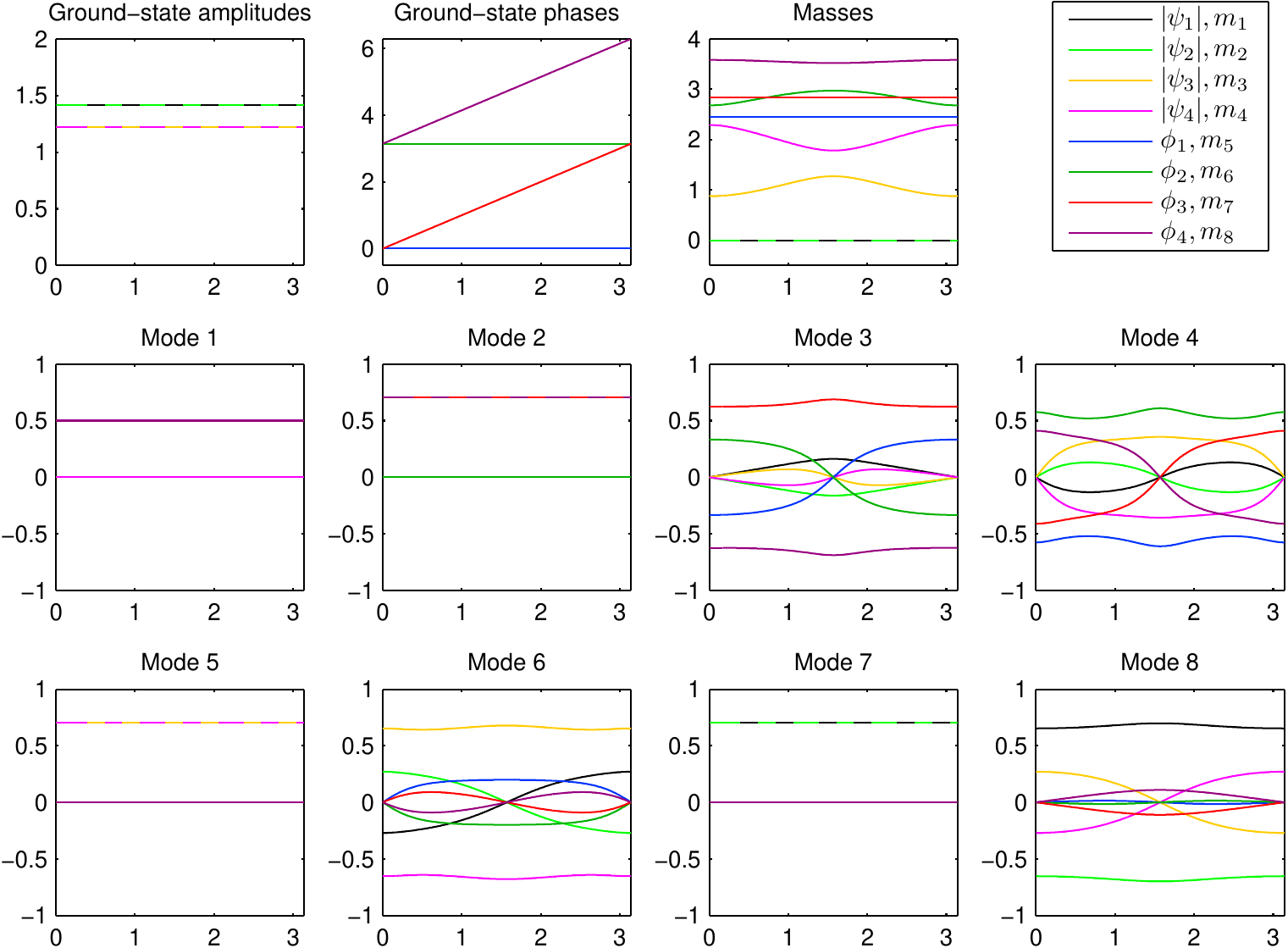}
  \caption{Ground states, (inverse) length scales, and normal modes with parameters as in Fig.~\ref{plot11_1} except that $\eta_{12} = -2$. The rotational angle $\gamma$ (Fig.~\ref{multfrust}) is plotted on the $x$-axes. Note that there is no third massless mode at the points $\gamma = 0, \pi$. Also, the modes corresponding to pure density excitations (modes 5 and 7) do not have the same length scales.}
  \label{plot11_2}
\end{figure*}

In the above, we have considered a multiply frustrated signature with maximal intercomponent symmetry. We now consider a case with less intercomponent symmetry. We choose the parameters to be as before, except that we change $\eta_{12}$ from $\eta_{12} = -1$ to $\eta_{12} = -2$. This has the effect of limiting the set of ground-state phase configurations. Whereas previously any configuration with pairwise antilocking was a ground-state configuration, the ground-state configurations are now those for which $\phi_1 = \phi_2 + \pi$ and $\phi_3 = \phi_4 + \pi$ (central circle in the left part of Fig.~\ref{symmetry}). Since $\phi_1$ and $\phi_2$ are now more strongly coupled, they are necessarily antilocked in the ground state. Consequently there is no longer any third massless mode, as can be seen from Fig.~\ref{plot11_2}, which displays ground states, normal modes and characteristic length scales for the case we now consider.

We have found that in the cases considered above, different ground states may be inequivalent in the sense of having quite different normal modes and characteristic length scales. This suggests that these new degeneracies do not correspond to broken symmetries. Also, the number of equivalent states is not the same for all ground states. Furthermore, note that in the case of complete intercomponent symmetry, the set of ground states does not form a manifold. (Too see this, consider the special points at which there exists a third massless mode.) Thus there can be no corresponding Lie group.

We note that for multiply frustrated signatures it is of no significance whether a particular coupling is attractive or repulsive; only the strengths of the couplings matter. This follows immediately from the fact that any multiply frustrated signature can be mapped to the signature for which all couplings are repulsive. This observation is also germane to the below, in which we consider frustrated three-component systems.

\subsection{Phase rotations: Other possibilities}
We are interested to know as generally as possible when energetically free phase rotations of the aforementioned type can occur. We begin by noting that phase frustration is required, and thus at least three components are required. The frustrated three-component signatures are all weakly equivalent; we choose the signature for which all Josephson couplings are repulsive. For this signature, one could imagine that two of the phases antilock due to strong repulsive coupling. See the central image in Fig.~\ref{multfrust} for an illustration (imagine that the two locked phases are one and the same). If the third phase is equally coupled to the two antilocked phases, then the third phase could rotate relative to the antilocked phases at no energy cost. We now show that this is not possible.

Requiring that the potential energy be stationary with respect to variations in the phase $\phi_1$, we obtain
\begin{equation*}
  - \eta_{12} |\psi_1| |\psi_2| \sin(\phi_1 - \phi_2)
  - \eta_{13} |\psi_1| |\psi_3| \sin(\phi_1 - \phi_3) = 0.
\end{equation*}
Assuming that $\phi_1$ and $\phi_2$ are antilocked, so that $\phi_1 - \phi_2 = \pi$, we find that the first term above is equal to zero. Thus the second term must also be equal to zero, whence $\phi_1$ and $\phi_3$ are either locked or antilocked. Thus there can be no energetically free phase rotations with fewer than four components.

The physical reason for the impossibility of energetically free phase rotations with only three components is clear: Firstly, antilocking is required, since without it the third phase will prefer certain values over others (recall that we assume that all couplings are repulsive; locking is an equivalent possibility for frustrated three-component signatures with attractive couplings). Assume, therefore, that $\phi_1$ and $\phi_2$ are antilocked, and envisage the insertion of $\phi_3$ so that $\phi_3 \neq \phi_i$ for $i \in \{1,2\}$. The couplings involving $\phi_3$ will cause $\phi_1$ and $\phi_2$ to be subjected to a net force, and thus antilocking will be broken. By the same argument, energetically free phase rotations are not possible for any singly frustrated signature.

\subsection{Condition for other continuous degeneracies}
We have studied multiply frustrated four-component systems with pairwise equivalence of phases in the ground states, and found that such systems can possess the aforementioned energetically free phase rotations. We now consider four-component systems with ground-state equivalence of two or three phases. Such systems can possess other types of continuous degeneracies.

We begin by considering the case of ground-state equivalence of two phases. (The case of ground-state equivalence of three phases is of course a special case of the case we now consider, as are the two cases previously considered.) We seek to investigate whether in the present case degeneracy can arise in the way described in Section~\ref{singdisc}, and if so under what conditions. To this end, we again parametrize the three relevant degrees of freedom in the phases as in the left half of Fig.~\ref{phaseparam}. The phases assumed to be equivalent are $\phi_1$ and $\phi_2$.

The part of the potential energy which depends on the phases is
\begin{multline}
  F = \tilde{\eta}_{12} \cos2\gamma
  + \tilde{\eta}_{34} \cos(\delta_1 + \delta_2) \\
  + \tilde{\eta}_{13} \cos(\delta_1 - \gamma)
  + \tilde{\eta}_{23} \cos(\delta_1 + \gamma) \\
  + \tilde{\eta}_{24} \cos(\delta_2 - \gamma)
  + \tilde{\eta}_{14} \cos(\delta_2 + \gamma),
  \label{totparamen}
\end{multline}
where we assume that $0 \leq \gamma \leq \pi/2$ (the other possibility $-\pi/2 \leq \gamma \leq 0$ is obtained by complex conjugation). In the following, $\delta_i$ and $\gamma$ are ground-state values unless the opposite is stated. By our assumption of ground-state equivalence of $\phi_1$ and $\phi_2$, we have that $\tilde{\eta}_{13} = \tilde{\eta}_{23} =: \tilde{\eta}_{1}$ and $\tilde{\eta}_{14} = \tilde{\eta}_{24} =: \tilde{\eta}_{2}$. Thus we can rewrite \eqref{totparamen} as
\begin{multline*}
  F = 2 (\tilde{\eta}_{1} \cos\delta_1 +
         \tilde{\eta}_{2} \cos\delta_2) \cos\gamma \\
      + \tilde{\eta}_{12} \cos 2\gamma
      + \tilde{\eta}_{34} \cos(\delta_1 + \delta_2).
\end{multline*}
Requiring that this energy be stationary with respect to variations of $\delta_i$ and $\gamma$, we obtain
\begin{align}
  \frac{\partial F}{\partial \delta_i} &=
    -2\tilde{\eta}_i \sin\delta_i \cos\gamma
    - \tilde{\eta}_{34} \sin(\delta_1 + \delta_2) = 0 \label{stateq1a} \\
  \frac{\partial F}{\partial \gamma} &=
    -2(\tilde{\eta}_{1} \cos\delta_1 +
       \tilde{\eta}_{2} \cos\delta_2) \sin\gamma
    - 2\tilde{\eta}_{12} \sin2\gamma = 0.
  \label{stateq1b}
\end{align}
As discussed in Sec.~\ref{singdisc}, it is only if $\gamma \neq 0$ that higher degeneracy can arise in the way described in that section. Therefore we assume that $\gamma \neq 0$; this happens precisely if $2\tilde{\eta}_{12} > |\tilde{\eta}_{1} \cos\delta_1 + \tilde{\eta}_{2} \cos\delta_2|$. Furthermore, it is easily seen that if $\gamma = \pi/2$ in a ground state, then there are energetically free phase rotations of the type described above. Therefore, we also assume that $\gamma \neq \pi/2$; this happens precisely if $\tilde{\eta}_{1} \cos\delta_1 + \tilde{\eta}_{2} \cos\delta_2 \neq 0$. In summary, we assume that $0 <  \gamma < \pi/2$, whence $0 < |\tilde{\eta}_{1} \cos\delta_1 + \tilde{\eta}_{2} \cos\delta_2| < 2\tilde{\eta}_{12}$.

Under the aforementioned assumptions, we have that \eqref{stateq1a} and \eqref{stateq1b} are equivalent to the following set of equations:
\begin{align}
  \label{stateq2a}
  0 &= \tilde{\eta}_1 \sin\delta_1 - \tilde{\eta}_2 \sin\delta_2 \\
  \label{stateq2b}
  0 &= 2\tilde{\eta}_1 \sin\delta_1 \cos\gamma
       + \tilde{\eta}_{34} \sin(\delta_1 + \delta_2) \\
  \label{stateq2c}
  0 &= 2\tilde{\eta}_{12} \cos\gamma
       + \tilde{\eta}_{1} \cos\delta_1 + \tilde{\eta}_{2} \cos\delta_2.
\end{align}
We note that the first equation above implies that $\delta_1 = 0 \mod \pi$ precisely if $\delta_2 = 0 \mod \pi$. This is of interest since if both $\delta_1 = 0 \mod \pi$ and $\delta_2 = 0 \mod \pi$, then the degeneracy corresponding to exchange of $\phi_1$ and $\phi_2$ will coincide with the degeneracy corresponding to complex conjugation. We thus assume that $\delta_1 \neq 0 \mod \pi$, whence $\delta_2 \neq 0 \mod \pi$.

We proceed by using \eqref{stateq2c} to substitute for $\cos\gamma$ in \eqref{totparamen}. (In doing so we use a trigonometric identity to substitute $\cos2\gamma$, so that the only remaining variables are the $\delta_i$.) The energy expression we thus obtain can, upon multiplying by a constant factor and disregarding an additive constant, be written as
\begin{equation*}
  G = -\tfrac{1}{2}
        (\tilde{\eta}_{1} \cos\delta_1 + \tilde{\eta}_{2} \cos\delta_2)^2
      + \tilde{\eta}_{12} \tilde{\eta}_{34} \cos(\delta_1 + \delta_2).
\end{equation*}
Considering $G$ amounts to restricting attention to a certain surface in phase space, which intersects all minima of interest. The variables $\delta_1$ and $\delta_2$ parametrize this surface. If, under our assumptions, $\delta_1$ and $\delta_2$ minimize $F$, then these same values will of course minimize $G$. We thus seek minima of $G$ by requiring stationarity with respect to variations of the $\delta_i$:
\begin{multline}
  \frac{\partial G}{\partial \delta_i} =
  (\tilde{\eta}_{1} \cos\delta_1 + \tilde{\eta}_{2} \cos\delta_2)
    \tilde{\eta}_i \sin\delta_i \\
  - \tilde{\eta}_{12} \tilde{\eta}_{34} \sin(\delta_1 + \delta_2) = 0.
  \label{stateq3}
\end{multline}
Using the assumption that $\delta_i \neq 0 \mod \pi$, as well as \eqref{stateq2a}, we can rewrite \eqref{stateq3} as
\begin{equation*}
  (\tilde{\eta}_1 \tilde{\eta}_2 - \tilde{\eta}_{12} \tilde{\eta}_{34})
  (\tilde{\eta}_{1} \cos\delta_1 + \tilde{\eta}_{2} \cos\delta_2) = 0.
\end{equation*}
We see that if $\tilde{\eta}_1 \tilde{\eta}_2 - \tilde{\eta}_{12} \tilde{\eta}_{34} \neq 0$ then $\tilde{\eta}_{1} \cos\delta_1 + \tilde{\eta}_{2} \cos\delta_2 = 0$, contradicting one of our assumptions. Thus it is only in cases for which
\begin{equation}
  \tilde{\eta}_1 \tilde{\eta}_2 - \tilde{\eta}_{12} \tilde{\eta}_{34} = 0
  \label{condition}
\end{equation}
that higher degeneracy can arise in the way described in Sec.~\ref{singdisc}, without there being energetically free phase rotations of the type investigated above. In fact, in some cases for which \eqref{condition} holds there \emph{is} additional \emph{continuous} degeneracy. One such case is that of complete intercomponent symmetry considered above; we consider other such cases below.

\begin{figure}
  \includegraphics{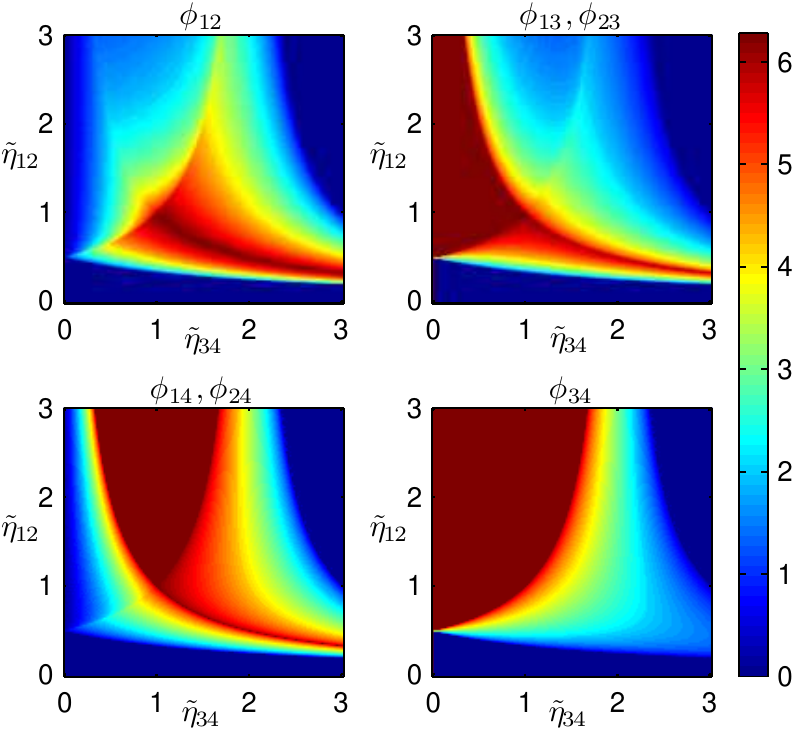}
  \caption{Sizes of the ranges over which ground-state phase differences can be varied for signature 11 under the assumption that $\phi_1$ and $\phi_2$ are equivalent in the ground states ($\tilde{\eta}_{13} = \tilde{\eta}_{23} =: \tilde{\eta}_{1}$ and $\tilde{\eta}_{14} = \tilde{\eta}_{24} =: \tilde{\eta}_{2}$), and that $\tilde{\eta}_1 \tilde{\eta}_2 - \tilde{\eta}_{12} \tilde{\eta}_{34} = 0$. We set $\tilde{\eta}_{2} = 1$. The curve $\tilde{\eta}_{12} \tilde{\eta}_{34} = 1$ coincides with the corresponding curve in Fig.~\ref{special1}.}
  \label{special2}
\end{figure}

\begin{figure*}
  \includegraphics{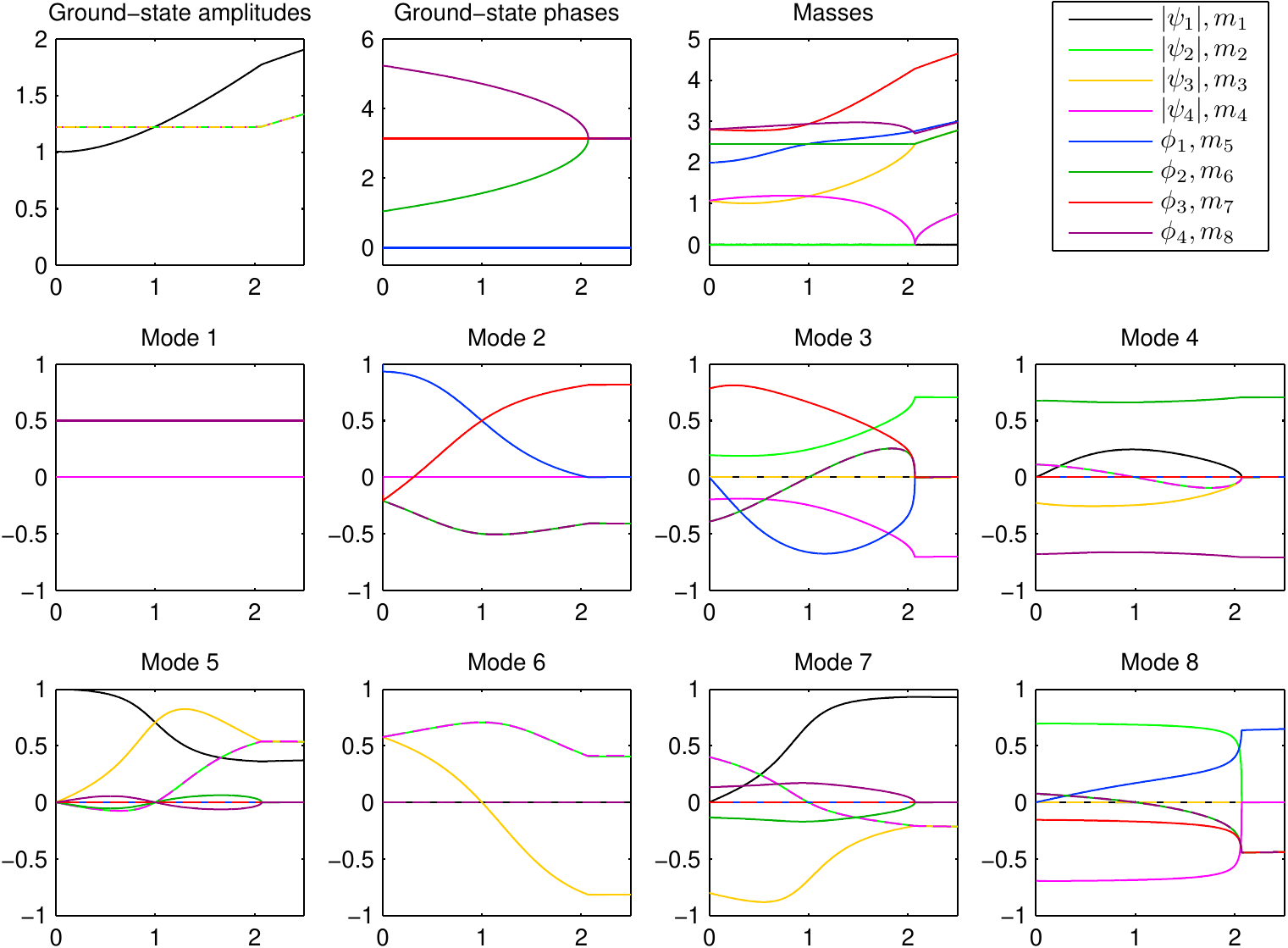}
  \caption{Ground states, (inverse) length scales, and normal modes for signature 11 with $\alpha_i = -1$, $\beta_i = 1$ and $\eta_{23} = \eta_{24} = \eta_{34} = -1$. The magnitude of the parameter $\eta := \eta_{12} = \eta_{13} = \eta_{14}$ is plotted on the $x$-axes. There are continuous ground-state degeneracies for a range of values of $\eta$. Note that for $\eta = -1$ the situation illustrated here is equivalent to that for $\gamma = \pi/2$ in Fig.~\ref{plot11_1}; for this value of $\eta$ there are energetically free phase rotations. For some $\eta \neq -1$ there are other types of continuous degeneracies; $\eta = -0.5$ corresponds to Fig.~\ref{spec1} and $\eta = -1.5$ corresponds to Fig.~\ref{spec2}.}
  \label{sixfold}
\end{figure*}

\begin{figure*}
  \includegraphics{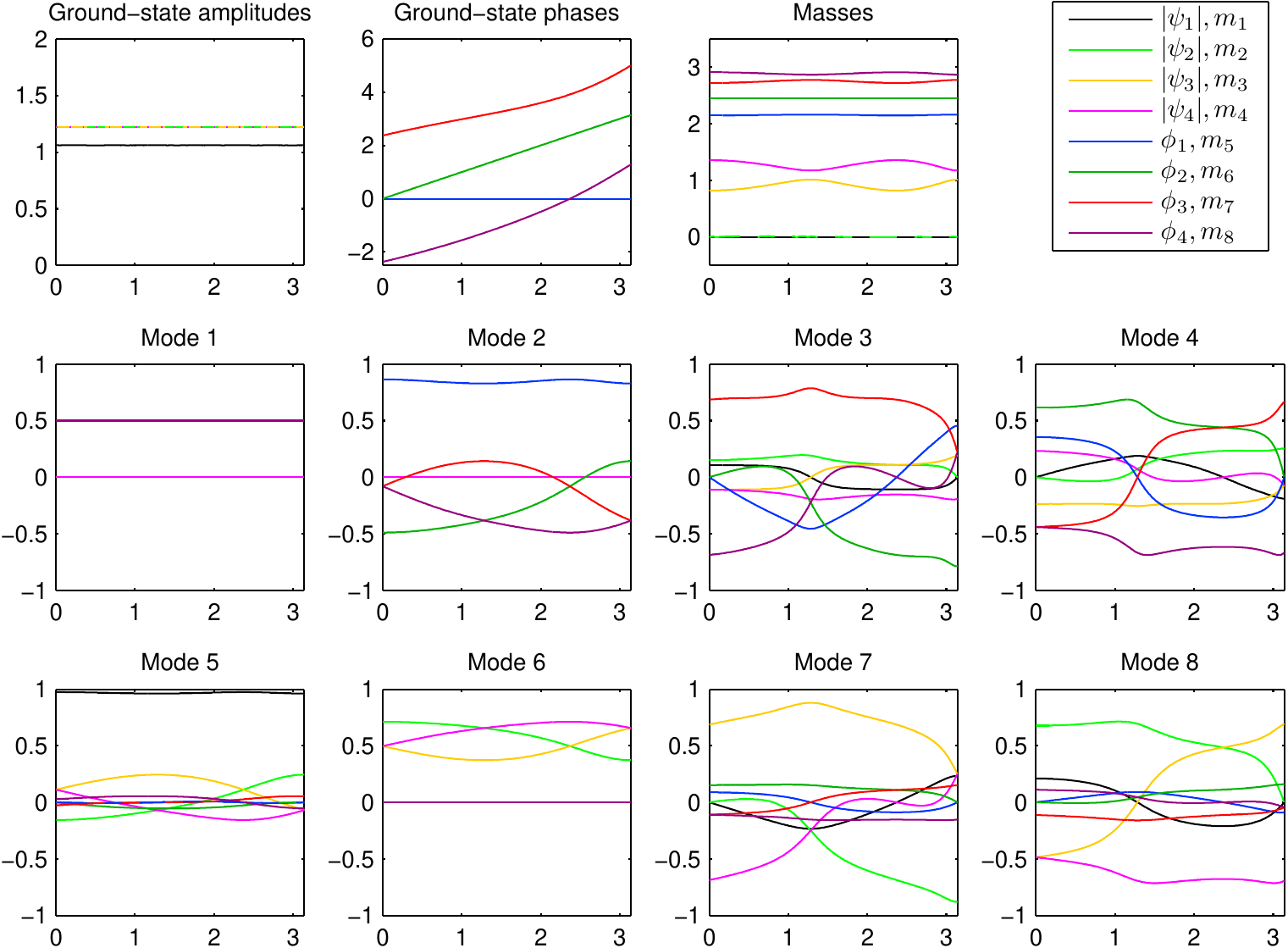}
  \caption{Ground states, (inverse) length scales, and normal modes with $\alpha_i = -1$, $\beta_i = 1$, $\eta_{23} = \eta_{24} = \eta_{34} = -1$ and $\eta := \eta_{12} = \eta_{13} = \eta_{14} = -0.5$ (cf.\ Fig.~\ref{sixfold}). The phase difference $-\phi_{12}$ is plotted on the $x$-axes. This phase difference can change by $2\pi$ without cost in potential energy. This can be seen from Fig.~\ref{special2} via either of the relabellings $1 \leftrightarrow 3$ or $1 \leftrightarrow 4$. Note that the state with $\phi_1 = \phi_2$ is equivalent to that with $\phi_1 = \phi_4$; thus we could have narrowed the displayed range of $\phi_{12}$ without loss of information.}
  \label{spec1}
\end{figure*}

\begin{figure*}
  \includegraphics{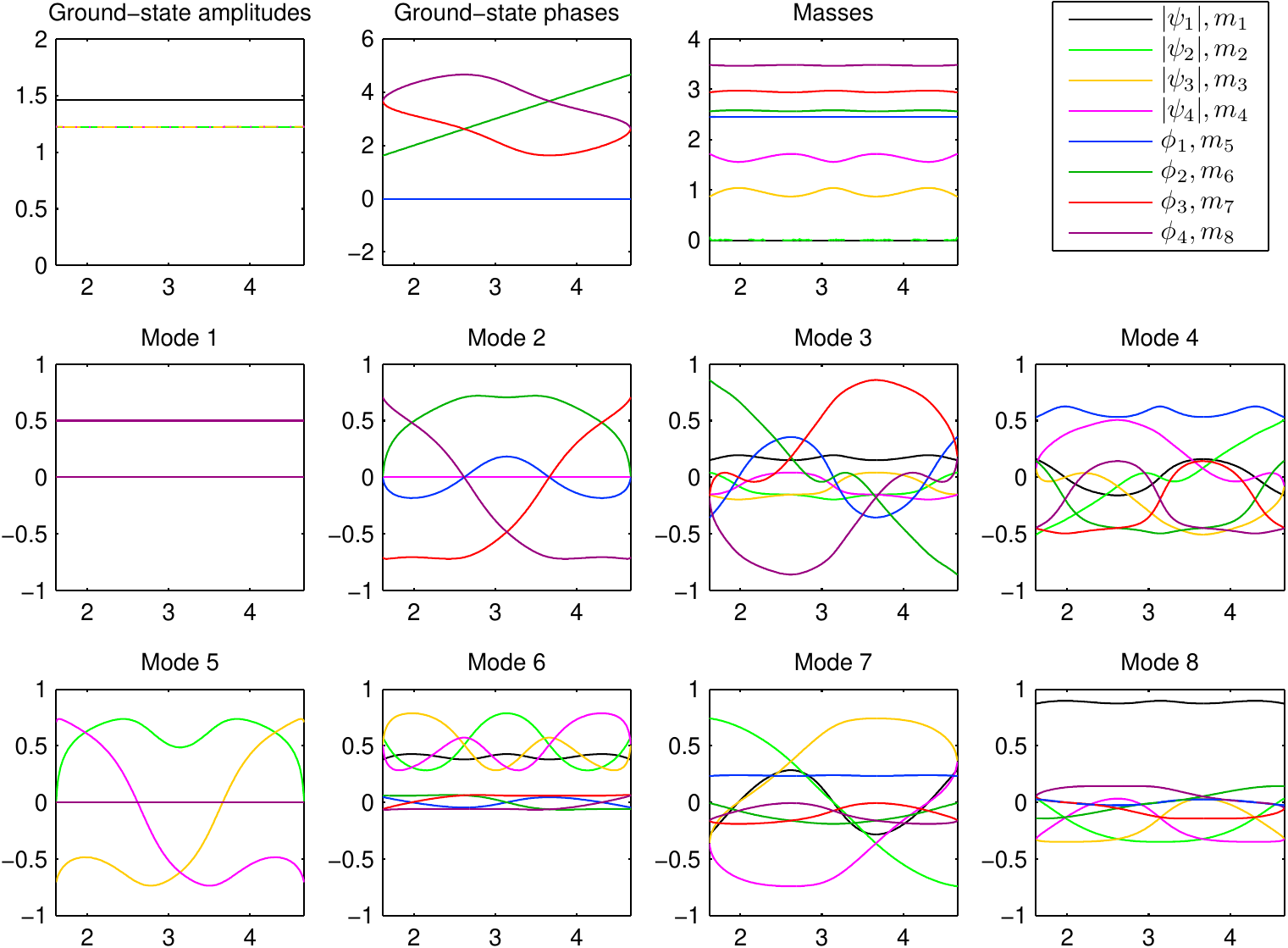}
  \caption{Ground states, (inverse) length scales, and normal modes with $\alpha_i = -1$, $\beta_i = 1$, $\eta_{23} = \eta_{24} = \eta_{34} = -1$ and $\eta := \eta_{12} = \eta_{13} = \eta_{14} = -1.5$ (cf.\ Fig.~\ref{sixfold}). The phase difference $-\phi_{12}$ is plotted on the $x$-axes. This phase difference can vary from $1.62$ to $4.67$ (the displayed range) without cost in potential energy. This can be seen (approximately) from Fig.~\ref{special2} via either of the relabellings $1 \leftrightarrow 3$ or $1 \leftrightarrow 4$. Note that, due to the equivalence of $\psi_2$, $\psi_3$ and $\psi_4$, the above graphs consist of three equivalent segments (e.g., the state with $\phi_2 = \phi_3$ is equivalent to that with $\phi_2 = \phi_4$).}
  \label{spec2}
\end{figure*}

The fact that continuous degeneracy can arise for parameter values that fulfil \eqref{condition} is made plausible by the following considerations. First, if \eqref{condition} holds then (under our assumptions) one can reduce \eqref{stateq2a}--\eqref{stateq2c} to \emph{two} equations. With only two equations for three variables, it is not surprising that one can have continuously connected degenerate minima of the potential energy. Second, consider the physical role of the (effective) coupling coefficients $\tilde{\eta}_1$, $\tilde{\eta}_2$, $\tilde{\eta}_{12}$ and $\tilde{\eta}_{34}$. The couplings corresponding to $\tilde{\eta}_1$ and $\tilde{\eta}_2$ together favour a phase configuration in which $\phi_1 = \phi_2 = \phi_3 + \pi = \phi_4 + \pi$. In contrast, the couplings corresponding to $\tilde{\eta}_{12}$ and $\tilde{\eta}_{34}$ favour a phase configuration in which $\phi_1 = \phi_2 + \pi$ and $\phi_3 = \phi_4 + \pi$. In the case of \eqref{condition}, there is a balance between these two tendencies which one could imagine gives rise to continuous ground-state degeneracy.

\subsection{Properties of other continuous degeneracies}
We now proceed to investigate the continuous ground-state degeneracies that can occur if \eqref{condition} is fulfilled [note that the assumption of ground-state equivalence of $\phi_1$ and $\phi_2$ is implicit in \eqref{condition}]. Without loss of generality, we set $\tilde{\eta}_{2} = 1$. This leaves us with two degrees of freedom in the $\tilde{\eta}_{ij}$. We choose to let these degrees of freedom be parametrized by $\tilde{\eta}_{12}$ and $\tilde{\eta}_{34}$; note that variation of these this implies variation of $\tilde{\eta}_{1}$. We consider the ranges $0 < \tilde{\eta}_{12}, \tilde{\eta}_{34} < 3$. For each corresponding point in the space of the $\tilde{\eta}_{ij}$, we determine the size of the range over which a given phase difference can be varied without leaving the set of ground states (Fig.~\ref{special2}). (We avoid the term ground-state manifold since the ground states do not necessarily form a manifold.) We see that there are regions with no degeneracy in phase differences, regions with complete degeneracy in certain phase differences, and regions with partial degeneracy in phase differences.

We now consider the example of three components being equivalent; this leads to fulfillment of \eqref{condition}, at least under appropriate relabeling. Let the free-energy parameters be such that $\alpha_i = -1$, $\beta_i = 1$ and $\eta_{23} = \eta_{24} = \eta_{34} = -1$. Consider various values of the parameters $\eta_{12} = \eta_{13} = \eta_{14}$; we assume that these are equal in order that $\psi_2$, $\psi_3$ and $\psi_4$ be equivalent. Also, we define $\eta$ to be the common value of $\eta_{12}$, $\eta_{13}$ and $\eta_{14}$. From Fig.~\ref{sixfold}, we can see that the ground-state values of the phases $\phi_2$, $\phi_3$ and $\phi_4$ are distinct for a range of values of $\eta$, and thus there is higher than twofold ground-state degeneracy. Note that for $\eta = -1$ the situation illustrated here is equivalent to that for $\gamma = \pi/2$ in Fig.~\ref{plot11_1}. For this particular value of $\eta$, the ground-state degeneracy corresponds to energetically free phase rotations. For other values of $\eta$ (investigated below), there are other kinds of continuous degeneracies. These degeneracies correspond to the second massless mode that is present for an entire range of values of $\eta$ (mode 2 in Fig.~\ref{sixfold}). This mode is a phase-only mode, which for $\eta = -1$ corresponds to energetically free phase rotations.

Consider the lines given by $\tilde{\eta}_{12} = 1$ in Fig.~\ref{special2}. It is easily seen that along these lines $\phi_1$, $\phi_2$ and $\phi_4$ are equivalent; we express this by saying that $\phi_3$ is \textit{the special phase}. Due to this equivalence of phases, we expect the lines corresponding to $\phi_{12}$ and $\phi_{14}$ to be equivalent; the same can be said of the lines corresponding to $\phi_{13}$ and $\phi_{34}$. Looking at Fig.~\ref{special2}, we see that this appears to be the case. In particular, we see that for $\tilde{\eta}_{34} \leq 1$ there is complete degeneracy in the phase differences $\phi_{13}$ and $\phi_{34}$, whereas for large values of $\tilde{\eta}_{34}$ there is no degeneracy in these phase differences. This is relevant for the cases studied in Fig.~\ref{sixfold}, in which $\phi_1$ is the special phase. Furthermore, it is easily seen that the same information can be obtained by considering the lines given by $\tilde{\eta}_{34} = 1$. In this case, it is $\phi_4$ that is the special phase. Also, the scale along these lines is inverted, so that taking the limit $\tilde{\eta}_{12} \rightarrow 0$ along these lines corresponds to taking the limit $\tilde{\eta}_{34} \rightarrow \infty$ along lines given by $\tilde{\eta}_{12} = 1$, and vice versa. Inspection of Fig.~\ref{special2} appears to confirm this.

We close this section by considering two more values of the parameter $\eta$ that is varied in Fig.~\ref{sixfold} (in addition to the value $\eta = -1$ already considered). We begin by considering the value $\eta = -0.5$ (Fig.~\ref{spec1}). On the basis of the discussion in the previous paragraph, we expect this value to give rise to complete ground-state degeneracy in the phase differences $\phi_{12}$, $\phi_{13}$ and $\phi_{14}$. This is indeed what we find. Thus, is may seem reasonable to say that there are energetically free phase rotations. However, note that here it is not the case that one set of phases can be rotated \emph{rigidly} relative to another at no energy cost. This is evident both from the plot of ground-state phases in Fig.~\ref{spec1}, and from the plot of the second massless mode in the same figure. The topology of the ground-state manifold is $[U(1)]^2 \times \mathbb{Z}_2$; this is the product of the broken symmetry $U(1) \times \mathbb{Z}_2$ and an additional factor of $U(1)$ stemming from accidental degeneracy.

We now consider the value $\eta = -1.5$ (Fig.~\ref{spec2}). By minimizing the potential energy for while keeping $\phi_{12}$ fixed, we find that $\phi_{12}$ can be varied in the range $1.62 \leq \phi_{12} \leq 4.67$ without leaving the ground-state manifold. By the equivalence of $\psi_{2}$, $\psi_{3}$ and $\psi_{4}$, this is also true of $\phi_{13}$ and $\phi_{14}$. Note that the ground-state degeneracy found here is roughly what one would expect on the basis of the results presented in Figs.~\ref{special2} and \ref{sixfold}. Finally, note that whereas the broken symmetry is $U(1) \times \mathbb{Z}_2$, the topology of the ground-state manifold is in fact $[U(1)]^2$. In other words, ground states that are related by the $\mathbb{Z}_2$ symmetry are in fact connected by accidental continuous degeneracy.

\section{Higher harmonics}

In the above, we have considered only first-harmonic Josephson couplings. In principle, there can also be higher harmonics of the form
\begin{equation}
  |\psi_i|^n |\psi_j|^n \cos n(\phi_i - \phi_j) \quad (n = 2,3,\dots).
  \label{higher}
\end{equation}
In general, the presence of such higher harmonics considerably complicates the situation. In particular, higher harmonics can give rise to metastability. In this section we make some observations pertaining to the case of second harmonics.

Consider chiral $p$-wave superconductors, which in certain cases can be modeled by a two-component model with biquadratic phase-coupling terms $(\psi_1\psi_2^*+\text{c.c.})^2$, corresponding to $n = 2$ in \eqref{higher} (see, e.g., Ref.~\onlinecite{agterberg1}). When expanding the potential to first order in fluctuations of the densities and phases around the ground state, the factor of $2$ in the second-harmonic Josephson coupling simply rescales the strength of the coupling, as compared to a first harmonic. Thus, the normal modes of two-band superconductors with either first-harmonic ($s$ or $s\pm$) or second-harmonic interband Josephson coupling are the same. This is true regardless of the signs of the couplings, and thus there are four equivalent possibilities.

From the above we conclude that mixing of phase and density modes is not a generic feature of systems with nontrivial (not $0$ or $\pi$) ground-state phase differences (in particular, it is not a consequence of TRSB). Rather, such mixing occurs when the cosine function in a Josephson coupling term is not stationary in the ground state, so that perturbation of the densities causes perturbation of the ground-state values of the phases, and vice versa. We now proceed to give what is perhaps the simplest example of this, in a system with only two components.

\begin{figure*}
  \includegraphics{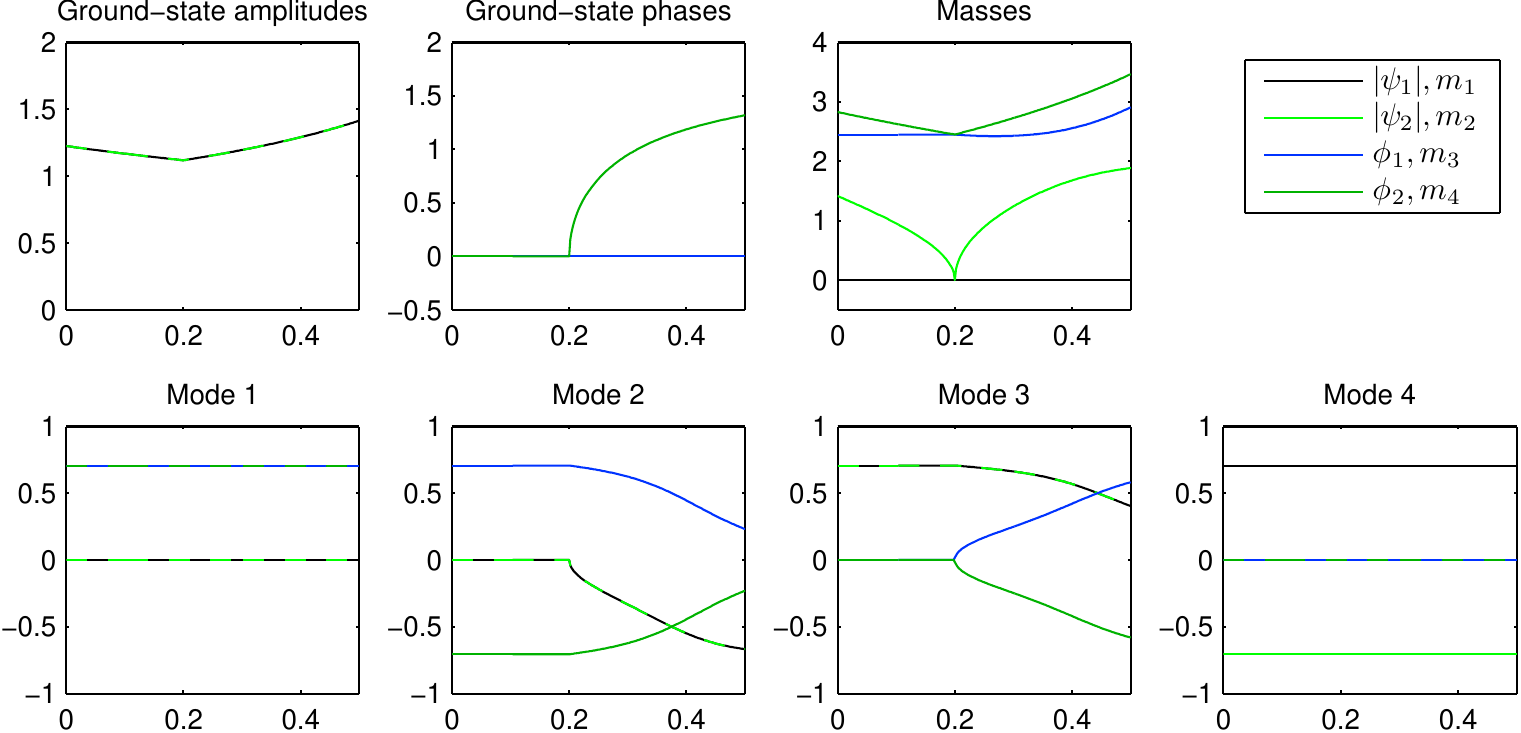}
  \caption{Ground states, (inverse) length scales, and normal modes with $\alpha_i = -1$, $\beta_i = 1$, and $\eta_{12} = 1$ in \eqref{ftcsh}. We plot $-\kappa_{12}$ on the $x$-axes. Note that for $\kappa_{12} < - 0.2$ there is phase-density mode mixing.}
  \label{twocomp}
\end{figure*}

Consider the following Ginzburg-Landau free-energy density:
\begin{multline}
  f = \tfrac{1}{2}(\nabla \times \mathbf{A})^2
      + \sum_{i=1}^2 \tfrac{1}{2} |\mathbf{D} \psi_i|^2
      + \alpha_i |\psi_i|^2 + \tfrac{1}{2}\beta_i |\psi_i|^4 \\
      - \eta_{12} |\psi_1| |\psi_2| \cos\phi_{12}
      - \kappa_{12} |\psi_1|^2 |\psi_2|^2 \cos2\phi_{12}.
  \label{ftcsh}
\end{multline}
But for the presence of the second-harmonic Josephson coupling, this is the same as \eqref{fgl} with $n = 2$. The inclusion of a second harmonic can lead to the ground state value of $\phi_{12}$ being nontrivial. We set $\eta_{12} = 1$ and vary $\kappa_{12}$ over the range $-0.5 < \kappa_{12} < 0$. For definiteness, we also set $\alpha_i = -1$ and $\beta_i = 1$. By straightforward extension of the calculations of Sec.~\ref{gsnm}, one can determine the normal modes and corresponding length scales for this system. The results are displayed in Fig.~\ref{twocomp}. We note that there is indeed phase-density mode mixing when Josephson coupling terms are not stationary in the ground state.

\section{Conclusion}
The abundance of recently discovered multiband superconductors with more than two bands (e.g., iron-based superconductors), and especially the possibility of creating Josephson-junction arrays using these materials, raises the need of also understanding more general multiband superconducting states. Here, we considered phenomenological Ginzburg-Landau and London models with an arbitrary number of superconducting components, with particular focus on four-component systems. It should be emphasized that realization of such systems does not require an intrinsically multiband system; rather, this can be achieved using real-space Josephson-coupled multilayers or Josephson-junction arrays.

We discussed the fact that the free energies \eqref{flondon} and \eqref{fgl} are invariant under the transformation which (i) inverts a given phase (i.e., adds $\pi$ to this phase), and (ii) changes the sign of all coupling coefficients involving this phase. Thus, apart from the trivial equivalence of signatures (corresponding to relabeling of components) which we call strong equivalence, there exists another equivalence of signatures (involving the aforementioned transformation), which we call weak equivalence.

We considered a graph-theoretical approach that allowed us to establish equivalence classes of multicomponent systems with frustrated intercomponent Josephson couplings. In this approach, we used a mapping where  each component corresponds to a vertex and each Josephson coupling corresponds to an edge. We thus found the following for an $n$-component system: The number of strong-equivalence classes is equal to the number of graphs on $n$ (unlabeled) vertices, and the number of weak-equivalence classes is equal to the number of switching classes of such graphs, i.e., the number of equivalence classes under Seidel switching.

We calculated ground states, normal modes and characteristic length scales for frustrated multicomponent superconductors modeled by the free-energy density \eqref{fgl}. We emphasize that the reported mixed phase-density modes are decoupled from the $U(1)$ sector. In the systems considered here, there can be fluctuation-driven phase transitions to anomalous normal states with broken discrete symmetry. The mixed modes can survive in these states due to the decoupling. For the case of four-component superconductors, we considered the two weak-equivalence classes of frustrated signatures that we call singly frustrated and multiply frustrated. We found that the properties of singly frustrated four-component systems are largely similar to those of phase-frustrated three-component systems.

In contrast, we found that multiply frustrated four-component signatures allow for qualitatively new features not present in the cases of two and three components. These are associated with \textit{accidental continuous ground-state degeneracies}, e.g.\ in the form of energetically free phase rotations. These degeneracies exist despite Josephson coupling between all phase pairs. More precisely, the degeneracies we have found can arise when at least two phases are equivalent in a ground state (in the weaker sense mentioned in Sec.~\ref{singdisc}). The existence of massless modes for some points in parameter space could lead to a number of interesting states for a range of parameters. For example, near such points in parameter space one of the coherence lengths can be anomalously large, leading to type-1.5 superconductivity.\cite{PRB.84.134518}

Note that the aforementioned continuous ground-state degeneracies, which do not correspond to spontaneously broken symmetries of the free energy, are such that the ground states do not all have the same length scales and normal modes. Furthermore, in the frustrated case of complete intercomponent symmetry, we found that the ground states do not form a manifold, whence there can be no corresponding Lie group.

Finally, we briefly considered systems with higher harmonics in the Josephson couplings. In doing so, we found that such systems typically display phase-density mode mixing, even in the simplest case of only two components. However, this is not the case for chiral $p$-wave superconductors;\cite{agterberg1} rather, these have the same normal modes as two-component $s$-wave superconductors.

In this paper we used an entirely phenomenological approach. Our results suggest that the case of four components is substantially richer than the better-investigated frustrated three-component case. This calls for further microscopic investigation of these new states based on approaches like those in, e.g., Refs.~\onlinecite{maiti,vafek}.

\begin{acknowledgments}
We thank Johan Carlstr\"om, Julien Garaud and Martin Speight for discussions. This work was supported by the Knut and Alice Wallenberg Foundation through a
Royal Swedish Academy of Sciences Fellowship, by the Swedish Research Council, and by the National Science Foundation CAREER Award No.~DMR-0955902. We thank the Swedish National Infrastructure for Computing (SNIC) at the National Supercomputer Center at Link\"oping, Sweden, for computational resources.
\end{acknowledgments}

%\bibliography{weston}

\begin{thebibliography}{39}%
\makeatletter
\providecommand \@ifxundefined [1]{%
 \@ifx{#1\undefined}
}%
\providecommand \@ifnum [1]{%
 \ifnum #1\expandafter \@firstoftwo
 \else \expandafter \@secondoftwo
 \fi
}%
\providecommand \@ifx [1]{%
 \ifx #1\expandafter \@firstoftwo
 \else \expandafter \@secondoftwo
 \fi
}%
\providecommand \natexlab [1]{#1}%
\providecommand \enquote  [1]{``#1''}%
\providecommand \bibnamefont  [1]{#1}%
\providecommand \bibfnamefont [1]{#1}%
\providecommand \citenamefont [1]{#1}%
\providecommand \href@noop [0]{\@secondoftwo}%
\providecommand \href [0]{\begingroup \@sanitize@url \@href}%
\providecommand \@href[1]{\@@startlink{#1}\@@href}%
\providecommand \@@href[1]{\endgroup#1\@@endlink}%
\providecommand \@sanitize@url [0]{\catcode `\\12\catcode `\$12\catcode
  `\&12\catcode `\#12\catcode `\^12\catcode `\_12\catcode `\%12\relax}%
\providecommand \@@startlink[1]{}%
\providecommand \@@endlink[0]{}%
\providecommand \url  [0]{\begingroup\@sanitize@url \@url }%
\providecommand \@url [1]{\endgroup\@href {#1}{\urlprefix }}%
\providecommand \urlprefix  [0]{URL }%
\providecommand \Eprint [0]{\href }%
\providecommand \doibase [0]{http://dx.doi.org/}%
\providecommand \selectlanguage [0]{\@gobble}%
\providecommand \bibinfo  [0]{\@secondoftwo}%
\providecommand \bibfield  [0]{\@secondoftwo}%
\providecommand \translation [1]{[#1]}%
\providecommand \BibitemOpen [0]{}%
\providecommand \bibitemStop [0]{}%
\providecommand \bibitemNoStop [0]{.\EOS\space}%
\providecommand \EOS [0]{\spacefactor3000\relax}%
\providecommand \BibitemShut  [1]{\csname bibitem#1\endcsname}%
\let\auto@bib@innerbib\@empty
%</preamble>
\bibitem [{\citenamefont {Moskalenko}(1959)}]{PMM.8.503}%
  \BibitemOpen
  \bibfield  {author} {\bibinfo {author} {\bibfnamefont {V.~A.}\ \bibnamefont
  {Moskalenko}},\ }\href@noop {} {\bibfield  {journal} {\bibinfo  {journal}
  {Phys. Met. Metallogr.}\ }\textbf {\bibinfo {volume} {8}},\ \bibinfo {pages}
  {503} (\bibinfo {year} {1959})}\BibitemShut {NoStop}%
\bibitem [{\citenamefont {Suhl}\ \emph {et~al.}(1959)\citenamefont {Suhl},
  \citenamefont {Matthias},\ and\ \citenamefont {Walker}}]{PRL.3.552}%
  \BibitemOpen
  \bibfield  {author} {\bibinfo {author} {\bibfnamefont {H.}~\bibnamefont
  {Suhl}}, \bibinfo {author} {\bibfnamefont {B.~T.}\ \bibnamefont {Matthias}},
  \ and\ \bibinfo {author} {\bibfnamefont {L.~R.}\ \bibnamefont {Walker}},\
  }\href {\doibase 10.1103/PhysRevLett.3.552} {\bibfield  {journal} {\bibinfo
  {journal} {Phys. Rev. Lett.}\ }\textbf {\bibinfo {volume} {3}},\ \bibinfo
  {pages} {552} (\bibinfo {year} {1959})}\BibitemShut {NoStop}%
\bibitem [{\citenamefont {Nagamatsu}\ \emph {et~al.}(2001)\citenamefont
  {Nagamatsu}, \citenamefont {Nakagawa}, \citenamefont {Muranaka},
  \citenamefont {Zenitani},\ and\ \citenamefont {Akimitsu}}]{Nature.410.63}%
  \BibitemOpen
  \bibfield  {author} {\bibinfo {author} {\bibfnamefont {J.}~\bibnamefont
  {Nagamatsu}}, \bibinfo {author} {\bibfnamefont {N.}~\bibnamefont {Nakagawa}},
  \bibinfo {author} {\bibfnamefont {T.}~\bibnamefont {Muranaka}}, \bibinfo
  {author} {\bibfnamefont {Y.}~\bibnamefont {Zenitani}}, \ and\ \bibinfo
  {author} {\bibfnamefont {J.}~\bibnamefont {Akimitsu}},\ }\href {\doibase
  10.1038/35065039} {\bibfield  {journal} {\bibinfo  {journal} {Nature}\
  }\textbf {\bibinfo {volume} {410}},\ \bibinfo {pages} {63} (\bibinfo {year}
  {2001})}\BibitemShut {NoStop}%
\bibitem [{\citenamefont {Xi}(2008)}]{RPP.71.116501}%
  \BibitemOpen
  \bibfield  {author} {\bibinfo {author} {\bibfnamefont {X.~X.}\ \bibnamefont
  {Xi}},\ }\href {\doibase 10.1088/0034-4885/71/11/116501} {\bibfield
  {journal} {\bibinfo  {journal} {Rep. Prog. Phys.}\ }\textbf {\bibinfo
  {volume} {71}},\ \bibinfo {pages} {116501} (\bibinfo {year}
  {2008})}\BibitemShut {NoStop}%
\bibitem [{\citenamefont {{Leggett}}(1966)}]{leggett}%
  \BibitemOpen
  \bibfield  {author} {\bibinfo {author} {\bibfnamefont {A.~J.}\ \bibnamefont
  {{Leggett}}},\ }\href {\doibase 10.1143/PTP.36.901} {\bibfield  {journal}
  {\bibinfo  {journal} {Progress of Theoretical Physics}\ }\textbf {\bibinfo
  {volume} {36}},\ \bibinfo {pages} {901} (\bibinfo {year} {1966})}\BibitemShut
  {NoStop}%
\bibitem [{\citenamefont {Blumberg}\ \emph {et~al.}(2007)\citenamefont
  {Blumberg}, \citenamefont {Mialitsin}, \citenamefont {Dennis}, \citenamefont
  {Klein}, \citenamefont {Zhigadlo},\ and\ \citenamefont
  {Karpinski}}]{PRL.99.227002}%
  \BibitemOpen
  \bibfield  {author} {\bibinfo {author} {\bibfnamefont {G.}~\bibnamefont
  {Blumberg}}, \bibinfo {author} {\bibfnamefont {A.}~\bibnamefont {Mialitsin}},
  \bibinfo {author} {\bibfnamefont {B.~S.}\ \bibnamefont {Dennis}}, \bibinfo
  {author} {\bibfnamefont {M.~V.}\ \bibnamefont {Klein}}, \bibinfo {author}
  {\bibfnamefont {N.~D.}\ \bibnamefont {Zhigadlo}}, \ and\ \bibinfo {author}
  {\bibfnamefont {J.}~\bibnamefont {Karpinski}},\ }\href {\doibase
  10.1103/PhysRevLett.99.227002} {\bibfield  {journal} {\bibinfo  {journal}
  {Phys. Rev. Lett.}\ }\textbf {\bibinfo {volume} {99}},\ \bibinfo {pages}
  {227002} (\bibinfo {year} {2007})}\BibitemShut {NoStop}%
\bibitem [{\citenamefont {Babaev}(2002)}]{frac}%
  \BibitemOpen
  \bibfield  {author} {\bibinfo {author} {\bibfnamefont {E.}~\bibnamefont
  {Babaev}},\ }\href {\doibase 10.1103/PhysRevLett.89.067001} {\bibfield
  {journal} {\bibinfo  {journal} {Phys. Rev. Lett.}\ }\textbf {\bibinfo
  {volume} {89}},\ \bibinfo {pages} {067001} (\bibinfo {year}
  {2002})}\BibitemShut {NoStop}%
\bibitem [{\citenamefont {Babaev}\ \emph {et~al.}(2010)\citenamefont {Babaev},
  \citenamefont {Carlstr\"om},\ and\ \citenamefont {Speight}}]{PRL.105.067003}%
  \BibitemOpen
  \bibfield  {author} {\bibinfo {author} {\bibfnamefont {E.}~\bibnamefont
  {Babaev}}, \bibinfo {author} {\bibfnamefont {J.}~\bibnamefont {Carlstr\"om}},
  \ and\ \bibinfo {author} {\bibfnamefont {M.}~\bibnamefont {Speight}},\ }\href
  {\doibase 10.1103/PhysRevLett.105.067003} {\bibfield  {journal} {\bibinfo
  {journal} {Phys. Rev. Lett.}\ }\textbf {\bibinfo {volume} {105}},\ \bibinfo
  {pages} {067003} (\bibinfo {year} {2010})}\BibitemShut {NoStop}%
\bibitem [{\citenamefont {Silaev}\ and\ \citenamefont
  {Babaev}(2011)}]{PRB.84.094515}%
  \BibitemOpen
  \bibfield  {author} {\bibinfo {author} {\bibfnamefont {M.}~\bibnamefont
  {Silaev}}\ and\ \bibinfo {author} {\bibfnamefont {E.}~\bibnamefont
  {Babaev}},\ }\href {\doibase 10.1103/PhysRevB.84.094515} {\bibfield
  {journal} {\bibinfo  {journal} {Phys. Rev. B}\ }\textbf {\bibinfo {volume}
  {84}},\ \bibinfo {pages} {094515} (\bibinfo {year} {2011})}\BibitemShut
  {NoStop}%
\bibitem [{\citenamefont {{Silaev}}\ and\ \citenamefont
  {{Babaev}}(2012)}]{silaevGL}%
  \BibitemOpen
  \bibfield  {author} {\bibinfo {author} {\bibfnamefont {M.}~\bibnamefont
  {{Silaev}}}\ and\ \bibinfo {author} {\bibfnamefont {E.}~\bibnamefont
  {{Babaev}}},\ }\href {\doibase 10.1103/PhysRevB.85.134514} {\bibfield
  {journal} {\bibinfo  {journal} {\prb}\ }\textbf {\bibinfo {volume} {85}},\
  \bibinfo {eid} {134514} (\bibinfo {year} {2012})}\BibitemShut {NoStop}%
\bibitem [{\citenamefont {Babaev}\ and\ \citenamefont
  {Speight}(2005)}]{PRB.72.180502}%
  \BibitemOpen
  \bibfield  {author} {\bibinfo {author} {\bibfnamefont {E.}~\bibnamefont
  {Babaev}}\ and\ \bibinfo {author} {\bibfnamefont {M.}~\bibnamefont
  {Speight}},\ }\href {\doibase 10.1103/PhysRevB.72.180502} {\bibfield
  {journal} {\bibinfo  {journal} {Phys. Rev. B}\ }\textbf {\bibinfo {volume}
  {72}},\ \bibinfo {pages} {180502} (\bibinfo {year} {2005})}\BibitemShut
  {NoStop}%
\bibitem [{\citenamefont {Carlstr\"om}\ \emph
  {et~al.}(2011{\natexlab{a}})\citenamefont {Carlstr\"om}, \citenamefont
  {Babaev},\ and\ \citenamefont {Speight}}]{PRB.83.174509}%
  \BibitemOpen
  \bibfield  {author} {\bibinfo {author} {\bibfnamefont {J.}~\bibnamefont
  {Carlstr\"om}}, \bibinfo {author} {\bibfnamefont {E.}~\bibnamefont {Babaev}},
  \ and\ \bibinfo {author} {\bibfnamefont {M.}~\bibnamefont {Speight}},\ }\href
  {\doibase 10.1103/PhysRevB.83.174509} {\bibfield  {journal} {\bibinfo
  {journal} {Phys. Rev. B}\ }\textbf {\bibinfo {volume} {83}},\ \bibinfo
  {pages} {174509} (\bibinfo {year} {2011}{\natexlab{a}})}\BibitemShut
  {NoStop}%
\bibitem [{\citenamefont {Carlstr\"om}\ \emph
  {et~al.}(2011{\natexlab{b}})\citenamefont {Carlstr\"om}, \citenamefont
  {Garaud},\ and\ \citenamefont {Babaev}}]{PRB.84.134518}%
  \BibitemOpen
  \bibfield  {author} {\bibinfo {author} {\bibfnamefont {J.}~\bibnamefont
  {Carlstr\"om}}, \bibinfo {author} {\bibfnamefont {J.}~\bibnamefont {Garaud}},
  \ and\ \bibinfo {author} {\bibfnamefont {E.}~\bibnamefont {Babaev}},\ }\href
  {\doibase 10.1103/PhysRevB.84.134518} {\bibfield  {journal} {\bibinfo
  {journal} {Phys. Rev. B}\ }\textbf {\bibinfo {volume} {84}},\ \bibinfo
  {pages} {134518} (\bibinfo {year} {2011}{\natexlab{b}})}\BibitemShut
  {NoStop}%
\bibitem [{\citenamefont {Geurts}\ \emph {et~al.}(2010)\citenamefont {Geurts},
  \citenamefont {Milo\ifmmode \check{s}\else
  \v{s}\fi{}evi\ifmmode~\acute{c}\else \'{c}\fi{}},\ and\ \citenamefont
  {Peeters}}]{PRB.81.214514}%
  \BibitemOpen
  \bibfield  {author} {\bibinfo {author} {\bibfnamefont {R.}~\bibnamefont
  {Geurts}}, \bibinfo {author} {\bibfnamefont {M.~V.}\ \bibnamefont
  {Milo\ifmmode \check{s}\else \v{s}\fi{}evi\ifmmode~\acute{c}\else
  \'{c}\fi{}}}, \ and\ \bibinfo {author} {\bibfnamefont {F.~M.}\ \bibnamefont
  {Peeters}},\ }\href {\doibase 10.1103/PhysRevB.81.214514} {\bibfield
  {journal} {\bibinfo  {journal} {Phys. Rev. B}\ }\textbf {\bibinfo {volume}
  {81}},\ \bibinfo {pages} {214514} (\bibinfo {year} {2010})}\BibitemShut
  {NoStop}%
\bibitem [{\citenamefont {Moshchalkov}\ \emph {et~al.}(2009)\citenamefont
  {Moshchalkov}, \citenamefont {Menghini}, \citenamefont {Nishio},
  \citenamefont {Chen}, \citenamefont {Silhanek}, \citenamefont {Dao},
  \citenamefont {Chibotaru}, \citenamefont {Zhigadlo},\ and\ \citenamefont
  {Karpinski}}]{PRL.102.117001}%
  \BibitemOpen
  \bibfield  {author} {\bibinfo {author} {\bibfnamefont {V.}~\bibnamefont
  {Moshchalkov}}, \bibinfo {author} {\bibfnamefont {M.}~\bibnamefont
  {Menghini}}, \bibinfo {author} {\bibfnamefont {T.}~\bibnamefont {Nishio}},
  \bibinfo {author} {\bibfnamefont {Q.~H.}\ \bibnamefont {Chen}}, \bibinfo
  {author} {\bibfnamefont {A.~V.}\ \bibnamefont {Silhanek}}, \bibinfo {author}
  {\bibfnamefont {V.~H.}\ \bibnamefont {Dao}}, \bibinfo {author} {\bibfnamefont
  {L.~F.}\ \bibnamefont {Chibotaru}}, \bibinfo {author} {\bibfnamefont {N.~D.}\
  \bibnamefont {Zhigadlo}}, \ and\ \bibinfo {author} {\bibfnamefont
  {J.}~\bibnamefont {Karpinski}},\ }\href {\doibase
  10.1103/PhysRevLett.102.117001} {\bibfield  {journal} {\bibinfo  {journal}
  {Phys. Rev. Lett.}\ }\textbf {\bibinfo {volume} {102}},\ \bibinfo {pages}
  {117001} (\bibinfo {year} {2009})}\BibitemShut {NoStop}%
\bibitem [{\citenamefont {Nishio}\ \emph {et~al.}(2010)\citenamefont {Nishio},
  \citenamefont {Dao}, \citenamefont {Chen}, \citenamefont {Chibotaru},
  \citenamefont {Kadowaki},\ and\ \citenamefont {Moshchalkov}}]{PRB.81.020506}%
  \BibitemOpen
  \bibfield  {author} {\bibinfo {author} {\bibfnamefont {T.}~\bibnamefont
  {Nishio}}, \bibinfo {author} {\bibfnamefont {V.~H.}\ \bibnamefont {Dao}},
  \bibinfo {author} {\bibfnamefont {Q.}~\bibnamefont {Chen}}, \bibinfo {author}
  {\bibfnamefont {L.~F.}\ \bibnamefont {Chibotaru}}, \bibinfo {author}
  {\bibfnamefont {K.}~\bibnamefont {Kadowaki}}, \ and\ \bibinfo {author}
  {\bibfnamefont {V.~V.}\ \bibnamefont {Moshchalkov}},\ }\href {\doibase
  10.1103/PhysRevB.81.020506} {\bibfield  {journal} {\bibinfo  {journal} {Phys.
  Rev. B}\ }\textbf {\bibinfo {volume} {81}},\ \bibinfo {pages} {020506}
  (\bibinfo {year} {2010})}\BibitemShut {NoStop}%
\bibitem [{\citenamefont {Dao}\ \emph {et~al.}(2011)\citenamefont {Dao},
  \citenamefont {Chibotaru}, \citenamefont {Nishio},\ and\ \citenamefont
  {Moshchalkov}}]{PRB.83.020503}%
  \BibitemOpen
  \bibfield  {author} {\bibinfo {author} {\bibfnamefont {V.~H.}\ \bibnamefont
  {Dao}}, \bibinfo {author} {\bibfnamefont {L.~F.}\ \bibnamefont {Chibotaru}},
  \bibinfo {author} {\bibfnamefont {T.}~\bibnamefont {Nishio}}, \ and\ \bibinfo
  {author} {\bibfnamefont {V.~V.}\ \bibnamefont {Moshchalkov}},\ }\href
  {\doibase 10.1103/PhysRevB.83.020503} {\bibfield  {journal} {\bibinfo
  {journal} {Phys. Rev. B}\ }\textbf {\bibinfo {volume} {83}},\ \bibinfo
  {pages} {020503} (\bibinfo {year} {2011})}\BibitemShut {NoStop}%
\bibitem [{\citenamefont {Babaev}\ \emph {et~al.}(2012)\citenamefont {Babaev},
  \citenamefont {Carlstr\"om}, \citenamefont {Garaud}, \citenamefont {Silaev},\
  and\ \citenamefont {Speight}}]{PhysicaC.479.2}%
  \BibitemOpen
  \bibfield  {author} {\bibinfo {author} {\bibfnamefont {E.}~\bibnamefont
  {Babaev}}, \bibinfo {author} {\bibfnamefont {J.}~\bibnamefont {Carlstr\"om}},
  \bibinfo {author} {\bibfnamefont {J.}~\bibnamefont {Garaud}}, \bibinfo
  {author} {\bibfnamefont {M.}~\bibnamefont {Silaev}}, \ and\ \bibinfo {author}
  {\bibfnamefont {J.~M.}\ \bibnamefont {Speight}},\ }\href {\doibase
  10.1016/j.physc.2012.01.002} {\bibfield  {journal} {\bibinfo  {journal}
  {Physica C}\ }\textbf {\bibinfo {volume} {479}},\ \bibinfo {pages} {2 }
  (\bibinfo {year} {2012})}\BibitemShut {NoStop}%
\bibitem [{\citenamefont {Kamihara}\ \emph {et~al.}(2008)\citenamefont
  {Kamihara}, \citenamefont {Watanabe}, \citenamefont {Hirano},\ and\
  \citenamefont {Hosono}}]{JACS.130.3296}%
  \BibitemOpen
  \bibfield  {author} {\bibinfo {author} {\bibfnamefont {Y.}~\bibnamefont
  {Kamihara}}, \bibinfo {author} {\bibfnamefont {T.}~\bibnamefont {Watanabe}},
  \bibinfo {author} {\bibfnamefont {M.}~\bibnamefont {Hirano}}, \ and\ \bibinfo
  {author} {\bibfnamefont {H.}~\bibnamefont {Hosono}},\ }\href {\doibase
  10.1021/ja800073m} {\bibfield  {journal} {\bibinfo  {journal} {J. Am. Chem.
  Soc.}\ }\textbf {\bibinfo {volume} {130}},\ \bibinfo {pages} {3296} (\bibinfo
  {year} {2008})}\BibitemShut {NoStop}%
\bibitem [{\citenamefont {Chu}\ \emph {et~al.}(2009)\citenamefont {Chu},
  \citenamefont {Koshelev}, \citenamefont {Kwok}, \citenamefont {Mazin},
  \citenamefont {Welp},\ and\ \citenamefont {Wen}}]{PhysicaC.469.313}%
  \BibitemOpen
  \bibfield  {author} {\bibinfo {author} {\bibfnamefont {P.~C.}\ \bibnamefont
  {Chu}}, \bibinfo {author} {\bibfnamefont {A.}~\bibnamefont {Koshelev}},
  \bibinfo {author} {\bibfnamefont {W.}~\bibnamefont {Kwok}}, \bibinfo {author}
  {\bibfnamefont {I.}~\bibnamefont {Mazin}}, \bibinfo {author} {\bibfnamefont
  {U.}~\bibnamefont {Welp}}, \ and\ \bibinfo {author} {\bibfnamefont {H.-H.}\
  \bibnamefont {Wen}},\ }\href {\doibase 10.1016/j.physc.2009.03.052}
  {\bibfield  {journal} {\bibinfo  {journal} {Physica C}\ }\textbf {\bibinfo
  {volume} {469}},\ \bibinfo {pages} {313 } (\bibinfo {year}
  {2009})}\BibitemShut {NoStop}%
\bibitem [{\citenamefont {Hirschfeld}\ \emph {et~al.}(2011)\citenamefont
  {Hirschfeld}, \citenamefont {Korshunov},\ and\ \citenamefont
  {Mazin}}]{RPP.74.124508}%
  \BibitemOpen
  \bibfield  {author} {\bibinfo {author} {\bibfnamefont {P.~J.}\ \bibnamefont
  {Hirschfeld}}, \bibinfo {author} {\bibfnamefont {M.~M.}\ \bibnamefont
  {Korshunov}}, \ and\ \bibinfo {author} {\bibfnamefont {I.~I.}\ \bibnamefont
  {Mazin}},\ }\href {\doibase 10.1088/0034-4885/74/12/124508} {\bibfield
  {journal} {\bibinfo  {journal} {Rep. Prog. Phys.}\ }\textbf {\bibinfo
  {volume} {74}},\ \bibinfo {pages} {124508} (\bibinfo {year}
  {2011})}\BibitemShut {NoStop}%
\bibitem [{\citenamefont {Ng}\ and\ \citenamefont
  {Nagaosa}(2009)}]{EPL.87.17003}%
  \BibitemOpen
  \bibfield  {author} {\bibinfo {author} {\bibfnamefont {T.~K.}\ \bibnamefont
  {Ng}}\ and\ \bibinfo {author} {\bibfnamefont {N.}~\bibnamefont {Nagaosa}},\
  }\href {\doibase 10.1209/0295-5075/87/17003} {\bibfield  {journal} {\bibinfo
  {journal} {Europhys. Lett.}\ }\textbf {\bibinfo {volume} {87}},\ \bibinfo
  {pages} {17003} (\bibinfo {year} {2009})}\BibitemShut {NoStop}%
\bibitem [{\citenamefont {Stanev}\ and\ \citenamefont {Te\ifmmode
  \check{s}\else \v{s}\fi{}anovi\ifmmode~\acute{c}\else
  \'{c}\fi{}}(2010)}]{PRB.81.134522}%
  \BibitemOpen
  \bibfield  {author} {\bibinfo {author} {\bibfnamefont {V.}~\bibnamefont
  {Stanev}}\ and\ \bibinfo {author} {\bibfnamefont {Z.}~\bibnamefont
  {Te\ifmmode \check{s}\else \v{s}\fi{}anovi\ifmmode~\acute{c}\else
  \'{c}\fi{}}},\ }\href {\doibase 10.1103/PhysRevB.81.134522} {\bibfield
  {journal} {\bibinfo  {journal} {Phys. Rev. B}\ }\textbf {\bibinfo {volume}
  {81}},\ \bibinfo {pages} {134522} (\bibinfo {year} {2010})}\BibitemShut
  {NoStop}%
\bibitem [{\citenamefont {{Maiti}}\ and\ \citenamefont
  {{Chubukov}}(2013)}]{maiti}%
  \BibitemOpen
  \bibfield  {author} {\bibinfo {author} {\bibfnamefont {S.}~\bibnamefont
  {{Maiti}}}\ and\ \bibinfo {author} {\bibfnamefont {A.~V.}\ \bibnamefont
  {{Chubukov}}},\ }\href {\doibase 10.1103/PhysRevB.87.144511} {\bibfield
  {journal} {\bibinfo  {journal} {\prb}\ }\textbf {\bibinfo {volume} {87}},\
  \bibinfo {eid} {144511} (\bibinfo {year} {2013})}\BibitemShut {NoStop}%
\bibitem [{\citenamefont {Mukherjee}\ and\ \citenamefont
  {Agterberg}(2011)}]{agterberg2011}%
  \BibitemOpen
  \bibfield  {author} {\bibinfo {author} {\bibfnamefont {S.}~\bibnamefont
  {Mukherjee}}\ and\ \bibinfo {author} {\bibfnamefont {D.~F.}\ \bibnamefont
  {Agterberg}},\ }\href {\doibase 10.1103/PhysRevB.84.134520} {\bibfield
  {journal} {\bibinfo  {journal} {Phys. Rev. B}\ }\textbf {\bibinfo {volume}
  {84}},\ \bibinfo {pages} {134520} (\bibinfo {year} {2011})}\BibitemShut
  {NoStop}%
\bibitem [{\citenamefont {{Mizushima}}\ \emph {et~al.}(2013)\citenamefont
  {{Mizushima}}, \citenamefont {{Takahashi}},\ and\ \citenamefont
  {{Machida}}}]{machidanew}%
  \BibitemOpen
  \bibfield  {author} {\bibinfo {author} {\bibfnamefont {T.}~\bibnamefont
  {{Mizushima}}}, \bibinfo {author} {\bibfnamefont {M.}~\bibnamefont
  {{Takahashi}}}, \ and\ \bibinfo {author} {\bibfnamefont {K.}~\bibnamefont
  {{Machida}}},\ }\href@noop {} {\bibfield  {journal} {\bibinfo  {journal}
  {ArXiv e-prints}\ } (\bibinfo {year} {2013})}\BibitemShut {NoStop}%
\bibitem [{\citenamefont {{Garaud}}\ \emph {et~al.}(2013)\citenamefont
  {{Garaud}}, \citenamefont {{Carlstr{\"o}m}}, \citenamefont {{Babaev}},\ and\
  \citenamefont {{Speight}}}]{cp22}%
  \BibitemOpen
  \bibfield  {author} {\bibinfo {author} {\bibfnamefont {J.}~\bibnamefont
  {{Garaud}}}, \bibinfo {author} {\bibfnamefont {J.}~\bibnamefont
  {{Carlstr{\"o}m}}}, \bibinfo {author} {\bibfnamefont {E.}~\bibnamefont
  {{Babaev}}}, \ and\ \bibinfo {author} {\bibfnamefont {M.}~\bibnamefont
  {{Speight}}},\ }\href {\doibase 10.1103/PhysRevB.87.014507} {\bibfield
  {journal} {\bibinfo  {journal} {\prb}\ }\textbf {\bibinfo {volume} {87}},\
  \bibinfo {eid} {014507} (\bibinfo {year} {2013})}\BibitemShut {NoStop}%
\bibitem [{\citenamefont {Lin}\ and\ \citenamefont
  {Hu}(2012)}]{PRL.108.177005}%
  \BibitemOpen
  \bibfield  {author} {\bibinfo {author} {\bibfnamefont {S.-Z.}\ \bibnamefont
  {Lin}}\ and\ \bibinfo {author} {\bibfnamefont {X.}~\bibnamefont {Hu}},\
  }\href {\doibase 10.1103/PhysRevLett.108.177005} {\bibfield  {journal}
  {\bibinfo  {journal} {Phys. Rev. Lett.}\ }\textbf {\bibinfo {volume} {108}},\
  \bibinfo {pages} {177005} (\bibinfo {year} {2012})}\BibitemShut {NoStop}%
\bibitem [{Note1()}]{Note1}%
  \BibitemOpen
  \bibinfo {note} {It has been argued that the transition to the TRSB phase is
  first order,\cite {PRB.81.134522} in which case there is no divergent length
  scale; however, a more recent paper\cite {maiti} supports the phase
  transition being second order.}\BibitemShut {Stop}%
\bibitem [{\citenamefont {{Stanev}}(2012)}]{stanev}%
  \BibitemOpen
  \bibfield  {author} {\bibinfo {author} {\bibfnamefont {V.}~\bibnamefont
  {{Stanev}}},\ }\href {\doibase 10.1103/PhysRevB.85.174520} {\bibfield
  {journal} {\bibinfo  {journal} {\prb}\ }\textbf {\bibinfo {volume} {85}},\
  \bibinfo {eid} {174520} (\bibinfo {year} {2012})}\BibitemShut {NoStop}%
\bibitem [{\citenamefont {{Garaud}}\ \emph {et~al.}(2011)\citenamefont
  {{Garaud}}, \citenamefont {{Carlstr{\"o}m}},\ and\ \citenamefont
  {{Babaev}}}]{cp21}%
  \BibitemOpen
  \bibfield  {author} {\bibinfo {author} {\bibfnamefont {J.}~\bibnamefont
  {{Garaud}}}, \bibinfo {author} {\bibfnamefont {J.}~\bibnamefont
  {{Carlstr{\"o}m}}}, \ and\ \bibinfo {author} {\bibfnamefont {E.}~\bibnamefont
  {{Babaev}}},\ }\href {\doibase 10.1103/PhysRevLett.107.197001} {\bibfield
  {journal} {\bibinfo  {journal} {Physical Review Letters}\ }\textbf {\bibinfo
  {volume} {107}},\ \bibinfo {eid} {197001} (\bibinfo {year}
  {2011})}\BibitemShut {NoStop}%
\bibitem [{\citenamefont {Lin}(2012)}]{PRB.86.014510}%
  \BibitemOpen
  \bibfield  {author} {\bibinfo {author} {\bibfnamefont {S.-Z.}\ \bibnamefont
  {Lin}},\ }\href {\doibase 10.1103/PhysRevB.86.014510} {\bibfield  {journal}
  {\bibinfo  {journal} {Phys. Rev. B}\ }\textbf {\bibinfo {volume} {86}},\
  \bibinfo {pages} {014510} (\bibinfo {year} {2012})}\BibitemShut {NoStop}%
\bibitem [{\citenamefont {{Smiseth}}\ \emph {et~al.}(2005)\citenamefont
  {{Smiseth}}, \citenamefont {{Sm{\o}rgrav}}, \citenamefont {{Babaev}},\ and\
  \citenamefont {{Sudb{\o}}}}]{smiseth}%
  \BibitemOpen
  \bibfield  {author} {\bibinfo {author} {\bibfnamefont {J.}~\bibnamefont
  {{Smiseth}}}, \bibinfo {author} {\bibfnamefont {E.}~\bibnamefont
  {{Sm{\o}rgrav}}}, \bibinfo {author} {\bibfnamefont {E.}~\bibnamefont
  {{Babaev}}}, \ and\ \bibinfo {author} {\bibfnamefont {A.}~\bibnamefont
  {{Sudb{\o}}}},\ }\href {\doibase 10.1103/PhysRevB.71.214509} {\bibfield
  {journal} {\bibinfo  {journal} {\prb}\ }\textbf {\bibinfo {volume} {71}},\
  \bibinfo {eid} {214509} (\bibinfo {year} {2005})}\BibitemShut {NoStop}%
\bibitem [{\citenamefont {{Bojesen}}\ \emph {et~al.}(2013)\citenamefont
  {{Bojesen}}, \citenamefont {{Babaev}},\ and\ \citenamefont
  {{Sudb{\o}}}}]{bojesen}%
  \BibitemOpen
  \bibfield  {author} {\bibinfo {author} {\bibfnamefont {T.}~\bibnamefont
  {{Bojesen}}}, \bibinfo {author} {\bibfnamefont {E.}~\bibnamefont {{Babaev}}},
  \ and\ \bibinfo {author} {\bibfnamefont {A.}~\bibnamefont {{Sudb{\o}}}},\
  }\href@noop {} {\bibfield  {journal} {\bibinfo  {journal} {ArXiv e-prints}\ }
  (\bibinfo {year} {2013})}\BibitemShut {NoStop}%
\bibitem [{\citenamefont {Harary}\ and\ \citenamefont
  {Palmer}(1973)}]{Harary.Palmer}%
  \BibitemOpen
  \bibfield  {author} {\bibinfo {author} {\bibfnamefont {F.}~\bibnamefont
  {Harary}}\ and\ \bibinfo {author} {\bibfnamefont {E.~M.}\ \bibnamefont
  {Palmer}},\ }\href@noop {} {\emph {\bibinfo {title} {Graphical
  Enumeration}}}\ (\bibinfo  {publisher} {Academic Press},\ \bibinfo {address}
  {New York},\ \bibinfo {year} {1973})\BibitemShut {NoStop}%
\bibitem [{\citenamefont {Mallows}\ and\ \citenamefont
  {Sloane}(1975)}]{SJAM.28.876}%
  \BibitemOpen
  \bibfield  {author} {\bibinfo {author} {\bibfnamefont {C.~L.}\ \bibnamefont
  {Mallows}}\ and\ \bibinfo {author} {\bibfnamefont {N.~J.~A.}\ \bibnamefont
  {Sloane}},\ }\href {http://www.jstor.org/stable/2100368} {\bibfield
  {journal} {\bibinfo  {journal} {SIAM J. Appl. Math.}\ }\textbf {\bibinfo
  {volume} {28}},\ \bibinfo {pages} {876} (\bibinfo {year} {1975})}\BibitemShut
  {NoStop}%
\bibitem [{\citenamefont {Anderson}(1963)}]{PR.130.439}%
  \BibitemOpen
  \bibfield  {author} {\bibinfo {author} {\bibfnamefont {P.~W.}\ \bibnamefont
  {Anderson}},\ }\href {\doibase 10.1103/PhysRev.130.439} {\bibfield  {journal}
  {\bibinfo  {journal} {Phys. Rev.}\ }\textbf {\bibinfo {volume} {130}},\
  \bibinfo {pages} {439} (\bibinfo {year} {1963})}\BibitemShut {NoStop}%
\bibitem [{\citenamefont {{Agterberg}}(1998)}]{agterberg1}%
  \BibitemOpen
  \bibfield  {author} {\bibinfo {author} {\bibfnamefont {D.~F.}\ \bibnamefont
  {{Agterberg}}},\ }\href {\doibase 10.1103/PhysRevLett.80.5184} {\bibfield
  {journal} {\bibinfo  {journal} {Physical Review Letters}\ }\textbf {\bibinfo
  {volume} {80}},\ \bibinfo {pages} {5184} (\bibinfo {year}
  {1998})}\BibitemShut {NoStop}%
\bibitem [{\citenamefont {{Cvetkovic}}\ and\ \citenamefont
  {{Vafek}}(2013)}]{vafek}%
  \BibitemOpen
  \bibfield  {author} {\bibinfo {author} {\bibfnamefont {V.}~\bibnamefont
  {{Cvetkovic}}}\ and\ \bibinfo {author} {\bibfnamefont {O.}~\bibnamefont
  {{Vafek}}},\ }\href@noop {} {\bibfield  {journal} {\bibinfo  {journal} {ArXiv
  e-prints}\ } (\bibinfo {year} {2013})},\ \Eprint
  {http://arxiv.org/abs/1304.3723} {arXiv:1304.3723 [cond-mat.str-el]}
  \BibitemShut {NoStop}%
\end{thebibliography}

\end{document}